\documentclass[11pt]{JHEP3}
\usepackage{amsmath,amssymb,bbm}   

\usepackage{graphicx}
\usepackage{psfrag}

\newcommand{\bR}{\mathbb{R}}

\newcommand{\bC}{\mathbb{C}}
\def\bt{\begin{thm}}
\def\et{\end{thm}}
\def\bp{\begin{prop}}
\def\ep{\end{prop}}
\def\bc{\begin{cor}}
\def\ec{\end{cor}}
\def\bl{\begin{lemma}}
\def\el{\end{lemma}}
\def\bd{\begin{dof}}
\def\ed{\end{dof}}

\def\square{\kern1pt\vbox
                 {\hrule height 0.6pt\hbox{\vrule width 0.6pt\hskip 3pt
      \vbox{\vskip 6pt}\hskip 3pt\vrule width 0.6pt}\hrule height 
0.6pt}\kern1pt}

\def\pf{\noindent{\it Proof:\ }}
\def\qed{\hfill\square}

\def\a{\alpha}

\def\d{\delta}

\def\o{\omega}

\def\G{\Gamma}

\def\O{\Omega}

\newfont{\goth}{eufm10 scaled \magstep1}

\font\sixrm=cmr6

\def\sc{\sixrm}

\def\n{\nabla}

\newtheorem{thm}{Theorem}
\newtheorem{prop}{Proposition}
\newtheorem{cor}{Corollary}
\newtheorem{lemma}{Lemma}
\newtheorem{dof}{Definition}

\def\ra{\rightarrow}

\def\be{\begin{equation}}
\def\ee{\end{equation}}

\def\re#1{(\ref{#1})}
\def\arr{\begin{array}{rlll}}
\def\ea{\end{array}}
\def\bea{\begin{eqnarray}}
\def\eea{\end{eqnarray}}

\newcommand{\mscr}[1]{\mbox{\scriptsize #1}}

\newcommand{\ft}[2]{{\textstyle\frac{#1}{#2}}}
\def\der{\partial}
\newcommand{\Imag} {{\Hat{\iota}}}
\newcommand{\id}   {{\mathbbm{1}}}

\newcommand{\refeq}[1]{(\ref{#1})}
\newcommand{\ih}{\Imag}
\newcommand{\overLine}[1]{\Bar{#1}}

\title{Special Geometry of Euclidean Supersymmetry II:\\
Hypermultiplets and the $c$-map}
\author{Vicente Cort\'es\\ Institut de Math\'ematiques \'Elie Cartan\\
  Universit\'e Henri Poincar\'e - Nancy I, 
  B.P.\ 239, F-54506 Vand{\oe}uvre-l\`es-Nancy, France\\
  cortes@iecn.u-nancy.fr
}
\preprint{FSU-TPI-02/05 \\ITP-05/06 \\ SPIN-05/04 }
\author{Christoph Mayer and Thomas Mohaupt\\
  Theoretisch-Physikalisches Institut \\
  Friedrich-Schiller-Universit\"{a}t Jena,
  Max-Wien-Platz 1,
  D-07743 Jena, Germany\\
  Christoph.Mayer,Thomas.Mohaupt@uni-jena.de
}
\author{Frank Saueressig\\
  Institute for Theoretical Physics and Spinoza Institute\\
  Postbus 80.195,
  3508 TD  Utrecht, The Netherlands\\
  F.S.Saueressig@phys.uu.nl
 }
\abstract{We construct two new versions of the $c$-map which allow 
us to obtain the target manifolds of hypermultiplets in Euclidean
theories with rigid ${\cal N} =2$ supersymmetry. While the 
Minkowskian para-$c$-map is obtained by dimensional reduction of
the Minkowskian vector multiplet lagrangian over time, the Euclidean
para-$c$-map corresponds to the dimensional reduction of the Euclidean
vector multiplet lagrangian. In both cases the resulting hypermultiplet
target spaces are para-hyper-K\"ahler manifolds. We review and prove 
the relevant results of para-complex and para-hypercomplex geometry.
In particular, we give a second, purely geometrical construction 
of both $c$-maps, by proving that the cotangent bundle $N=T^*M$ of any
affine special (para-)K\"ahler manifold $M$ is para-hyper-K\"ahler.}
\begin{document}
\section{Introduction, summary and outlook}

\subsection{Introduction}

In this paper we continue the investigation of the geometrical 
structures of Euclidean supersymmetric theories initiated in 
\cite{CMMS}. As explained there in more detail, the main motivation
of our work is to obtain a better understanding of instantons in
supersymmetric
string theory compactifications and their rigidly supersymmetric field
theory limits. Moreover, since solitons and instantons are mutually
related by dimensional reduction over time, instantons can be used
as the starting point for the systematic construction of solitonic 
solutions, such as black holes, black branes and domain walls \cite{Oxidation}.
One 
also expects to be able to generate cosmological solutions of 
the type II${}^*$ string theories, which are related to the standard
type II string theories by time-like T-duality \cite{Hull,BehrndtCvetic}. 
Since the target space geometries of the scalar sigma models appearing in 
Euclidean supersymmetric theories differ from those of the Minkowskian
theories \cite{Schwinger,zumino,vN,blau,BelVanVan,CreEtAl,HulJul}, 
we were lead to the question what characterizes these
geometries. For theories with more than 8 supercharges the 
scalar geometry is completely fixed by the matter content, while 
for theories with 8 or less supercharges the geometry is restricted, but
not unique. The boundary case of 8 supercharges is particularly interesting,
because the resulting theories have a rich structure while they (or at
least certain aspects of them) can be treated exactly. For space-times
with Minkowskian signature, the corresponding geometries
are known as the `special geometries.'\footnote{We refer to 
\cite{VanPro} for a review.} 
More precisely,
for locally supersymmetric theories
the manifolds spanned by the scalars of vector multiplets 
are `projective very special real  manifolds' \cite{GST,deWvP,ACDV}
in five dimensions
and `projective special K\"ahler manifolds' \cite{DLP,Strominger,Freed,ACD} 
in four dimensions,
while the scalar manifolds of hypermultiplets are 
quaternion-K\"ahler \cite{BagWit,DDKV}. 
For rigidly
supersymmetric theories the manifolds spanned by the scalars of
vector multiplets are `affine very special real manifolds' \cite{CMMS} in five
dimensions and `affine special K\"ahler manifolds' \cite{Gates,CRTV,Freed,ACD} in four 
dimensions,
while the scalar manifolds of hypermultiplets are hyper-K\"ahler manifolds 
\cite{AlvGauFre} in all dimensions $\leq 6$.

In \cite{CMMS} we investigated vector multiplets
in Euclidean four-dimensional space. We introduced the notion of an
`affine special para-K\"ahler manifold' and showed that these are the target
manifolds of rigid Euclidean vector multiplets. The basic difference
between the scalar geometries of Minkowskian and Euclidean vector multiplets
is  that the complex structure $I$, $I^2 = - \mathbbm{1}$ is replaced 
by a para-complex structure $J$, $J^2 = \mathbbm{1}$ (with equal
rank of the eigenspaces for the eigenvalues $\pm 1$). As discussed 
in \cite{CMMS}, all the relevant notions of complex geometry, 
such has Hermitian, K\"ahler and special K\"ahler, have their precise
analogues in para-complex geometry.

The purpose of this paper is to identify the scalar geometry of
Euclidean hypermultiplets in rigidly supersymmetric theories. Since
the scalar geometry of hypermultiplets does not change under dimensional
reduction, this geometry is the same for all dimensions where 
hypermultiplets exist, i.e., for dimensions $\leq 6$. Our main result
is that the scalar manifolds of Euclidean hypermultiplets are
para-hyper-K\"ahler manifolds. The precise definition will be
given later, but let us already characterize them heuristically as
para-complex analogues of hyper-K\"ahler manifolds:
hyper-K\"ahler manifolds have three
complex structures, which satisfy the 
quaternionic algebra under multiplication, while para-hyper-K\"ahler 
manifolds have 
two para-complex structures and one complex structure, which satisfy
the so-called para-quaternionic algebra. Thus the resulting picture
is very similar to the one we found for vector multiplets.

The most convenient
tool to find these Euclidean hypermultiplet manifolds is to construct
a new version of the so-called $c$-map, which is well known from 
Minkowskian theories \cite{CFG,FS,deWvP,DDKV}. 
In fact, we will construct {\em two} new $c$-maps,
called the Minkowskian and the Euclidean para-$c$-map (for reasons 
that will become obvious in a moment). Each of the $c$-maps is constructed 
using two complementary
approaches, a physical and a mathematical one. While the physical
approach, based on T-duality and dimensional reduction of lagrangians, 
provides an explicit expression for the hypermultiplet metric, the 
geometrical structures of the hypermultiplet manifolds are not manifest
and their identification requires a lot of work. The mathematical 
approach provides a geometrical construction of the hypermultiplet
manifolds and the origin of all its structures is transparent and
manifest. When formulated in terms of coordinates which
correspond to the physical scalar fields, it is straightforward to
verify that the metrics obtained in both constructions are identical.

The outline of this paper is as follows: in the reminder of this
secion we give a comprehensive summary and discussion of the results,
and we also give a brief outlook on future directions of research. 
Section 2 studies the $c$-maps from the mathematical point of view.
Here we provide the definitions for all the relevant notions of
para-complex and para-hypercomplex geometry, and we formulate
and prove five theorems which specify the properties of 
the $c$-maps. Section 3 gives the physical
treatment of the $c$-maps through the dimensional reduction of
supersymmetric lagrangians from four
to three dimensions. The resulting hypermultiplet metric is shown
to be of the type constructed in Section 2. In Section 4 we 
discuss a restricted class of hypermultiplet manifolds which can
be obtained by dimensional reduction from five to three dimensions.
The reduction from five to four dimensions is related to 
the so-called $r$-maps, and we can summarize the relations between
$r$-maps and $c$-maps in a commutative diagram.

\subsection{Summary and discussion}

The physical way to understand the $c$-map is its
relation to T-duality in string theory. To be concrete let us
consider type-IIA and type-IIB string theory, both in a space-time
background of the form $M_d \times X_{10-d}$, where $M_d$ is 
$d$-dimensional Minkowski space and $X_{10-d}$ is a $(10-d)$-dimensional
Ricci-flat compact manifold. Then T-duality states that type-IIA string theory
in the background $M_{d-1} \times S_1^R \times X_{10 - d}$ is identical to 
type-IIB string theory in the background 
$M_{d-1} \times S_1^{R^{-1}} \times X_{10-d}$.
Here $M_{d-1}$ is $(d-1)$-dimensional Minkowski space,
$S_1^R$ is a circle of radius $R$, and $S_1^{R^{-1}}$ is a circle with
the inverse radius. (The radius is measured in terms of the fundamental
length scale $\sqrt{\alpha'}$ of string theory.) In other words the 
dimensional reduction from $d$ to $d-1$ dimensions yields {\em one} continuous
family of theories, labelled by $R$, which has two distinct $d$-dimensional
limits $R \rightarrow \infty$ (IIA theory on $M_d \times X_{10-d}$) and
$R^{-1} \rightarrow \infty$ (IIB theory on $M_d \times X_{10-d}$):
\[
\mbox{IIA} / X_{10-d} \leftarrow \mbox{IIA} / ( X_{10-d} \times S_1^R ) = 
\mbox{IIB} / (X_{10-d} \times S_1^{R^{-1}}) \rightarrow 
\mbox{IIB} / X_{10-d} \;.
\]
To obtain the $c$-map it is sufficient to go from the full string theories
to the $d$-dimensional effective supergravity theories which describe their
massless modes. We now choose  $d=4$
and take $X$ to be a (generic) Calabi-Yau threefold $X_6$. Then one
has two four-dimensional supergravity theories with ${\cal N}=2$
supersymmetry (8 supercharges) which are related as follows:
given one of the two supergravity actions, one performs a dimensional
reduction on a circle of radius $R$ and obtains a three-dimensional
supergravity action. Then one performs the limit $R^{-1} \rightarrow \infty$
and re-interprets the result as a four-dimensional theory. In practise,
this can be done by finding a four-dimensional theory which reproduces
the three-dimensional theory upon dimensional reduction on a circle
of radius $R^{-1}$. In the case at hand, this construction maps
the vector multiplets (hypermultiplets) of the four-dimensional type-IIA
theory to the hypermultiplets (vector multiplets) of the
four-dimensional type-IIB theory. Recall that for Minkowski signature
a four-dimensional
vector multiplet $(A_\mu, \lambda_i, z)$ consists of a gauge field
$A_\mu$, a doublet of Majorana spinors $\lambda_i$, $i=1,2$ and 
a complex scalar $z$. Under dimensional reduction the vector 
$A_\mu$ decomposes into a scalar $A_3$ and a vector $A_m$, $m=0,1,2$.
Moreover, a three-dimensional vector field can be dualized into 
a scalar field,\footnote{See Section 3 for the details.}
so that all together we obtain a three-dimensional hypermultiplet
which consists of four real scalars and a doublet of Majorana
spinors. In the limit $R^{-1} \rightarrow \infty$ this becomes
a four-dimensional hypermultiplet with the same scalar manifold. 
This defines a map, known as the $c$-map, 
between the vector multiplet manifold $M$
of the four-dimensional IIA theory and the hypermultiplet manifold
$N$  of the four-dimensional IIB theory. Note that the $c$-map is
already determined by the dimensional reduction of the bosonic part of 
the vector multiplet lagrangian from four to three dimensions: 
\[
{\small 
c: M = \{ \mbox{mfd. of 4d vector multiplet scalars} \} \rightarrow 
N = \{ \mbox{mfd. of 3d/4d hypermultiplet scalars} \} \;.}
\]
The opposite happens for the hypermultiplets of the IIA theory. 
Under dimensional reduction they become three-dimensional hypermultiplets, 
but in order to perform the limit $R^{-1} \rightarrow \infty$ one
needs to dualize one of the scalars into a three-dimensional vector,
which becomes a four-dimensional vector
in the decompactification limit.
This way the hypermultiplets of the IIA theory are mapped to the
vector multiplets of the IIB theory. The resulting map between the 
scalar manifolds is the inverse $c$-map.

So far we discussed the $c$-map in the context of supergravity, which
is natural because string theory automatically incorporates gravity.
But one can also consider a limit where gravity decouples. Then one 
obtains a $c$-map, called the rigid $c$-map,\footnote{We will omit  
the `rigid' in the following.} 
between the vector and hypermultiplet manifolds of
rigidly supersymmetric theories. 
In four-dimensional rigid 
${\cal N}=2$ superymmetry, the target space $M$
of the vector multiplet scalars must be an affine special 
K\"ahler manifold of real dimension $2n$, where $n$ is the number
of vector multiplets. Since hypermultiplet manifolds must be
hyper-K\"ahler, the $c$-map assigns to every affine special 
K\"ahler manifold $M$ a hyper-K\"ahler manifold $N$ of twice the dimension.
Note, however, that not all hyper-K\"ahler manifolds can be obtained
this way. In fact, the hyper-K\"ahler manifolds in the image
of $c$-map are non-generic, because they have isometries originating
from the gauge symmetries of the vector fields which
have been dualized into scalars.

In this paper we will use two modified $c$-maps to construct Euclidean
hypermultiplet manifolds from vector multiplet manifolds. One way is
to start from the vector multiplet lagrangians of Euclidean four-dimensional
theories, which were  constructed in \cite{CMMS}, 
and to reduce the theory to 
three dimensions. Again the four-dimensional vector decomposes into a
three-dimensional vector and a scalar, and the vector can be dualized
into another scalar. This way we obtain a three-dimensional Euclidean 
hypermultiplet lagrangian, from which we can read off the 
hypermultiplet manifold. The resulting $c$-map is called 
the Euclidean para-$c$-map, denoted $c^{4+0}_{3+0}$. While the initial
vector multiplet manifold $M$ is affine special para-K\"ahler,
the resulting hypermultiplet manifold $N$ is para-hyper-K\"ahler:
\be
c^{4+0}_{3+0} : \{ \mbox{affine special para-K\"ahler mfds.}\} \rightarrow
\{ \mbox{para-hyper-K\"ahler mfds.} \} \;.
\ee

The second option that we have to construct Euclidean hypermultiplet
manifolds is to start with a Minkowskian vector multiplet lagrangian
and to perform a dimensional reduction over time. This defines
a map $c^{3+1}_{3+0}$, called the Minkowskian para-$c$-map 
which assigns to every Minkowskian 
vector multiplet manifold a Euclidean hypermultiplet manifold. 
While the Minkowskian vector multiplet manifolds are special 
K\"ahler manifolds, the Euclidean hypermultiplet
manifolds are again para-hyper-K\"ahler manifolds:
\be
c^{3+1}_{3+0} : \{ \mbox{affine special K\"ahler mfds.}\} \rightarrow
\{ \mbox{para-hyper-K\"ahler mfds.} \} \;.
\ee

\begin{figure*}[t]
  \begin{center}
    \psfrag{r4+13+1}[][][1.2361]{$r^{4+1}_{3+1}$}
    \psfrag{r4+14+0}[][][1.2361]{$r^{4+1}_{4+0}$}
    \psfrag{c3+13+0}[][][1.2361]{$c^{3+1}_{3+0}$}
    \psfrag{c4+03+0}[][][1.2361]{$c^{4+0}_{3+0}$}
    \psfrag{q}[][][1.2361]{$q^{4+1}_{3+0}$}
    \psfrag{VSR}[][][1]{very special real}
    \psfrag{SK}[][][1]{special K\"ahler}
    \psfrag{PSK}[][][1]{special para-K\"ahler}
    \psfrag{PHK}[][][1]{special para-hyper-K\"ahler}
    \includegraphics[width=\textwidth]{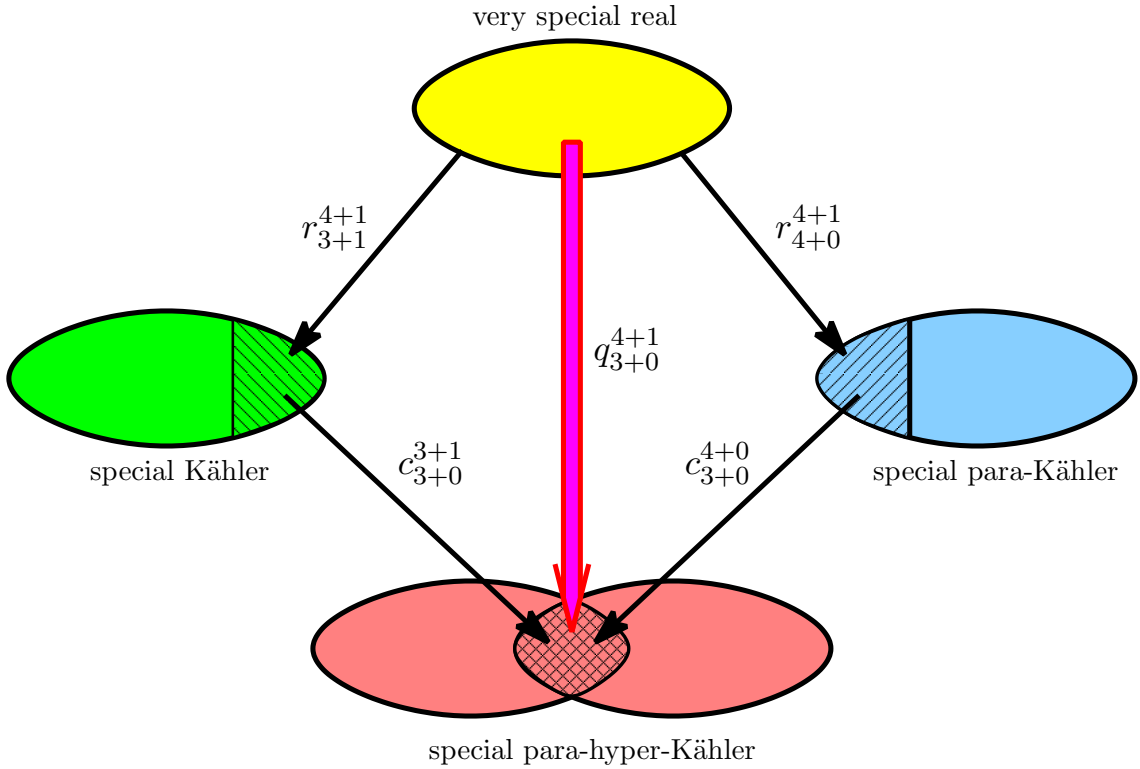}
  \end{center}
  \caption{\label{Diagram} This figure summarizes how the scalar manifolds of 
three-dimensional Euclidean hypermultiplets (at the bottom) 
are related to those of
Minkowskian and Euclidean four-dimensional vector multiplets (in the middle)
and of five-dimensional Minkowskian vector multiplets (at the top). 
Dimensional
reduction does only yield a special subset of the lower dimensional theories:
the scalars of five-dimensional vector multiplets parametrize a very special
real manifold, which is encoded in a real polynomial of degree 3, the
prepotential. The reduction over 
space (time) 
gives very special (para-)K\"ahler manifolds (hatched segments of 
left and right middle blob, respectively). 
These are determined by the real
cubic prepotential of the five-dimensional theory, while generic special 
(para-)K\"ahler manifolds have a general
(para-)holomorphic prepotential (full blobs in the middle). 
The reduction of general four-dimensional 
Minkowskian (Euclidean) vector multiplets over time (space) gives special 
para-hyper-K\"ahler manifolds (intersecting blobs at the bottom), which 
are determined by the (para-)holomorphic
prepotential of (at least) one of the four-dimensional theories. 
The reduction from five to 
three dimensions yields  very special para-hyper-K\"ahler manifolds
(doubly-hatched segment of the lower blobs), which are 
encoded in a real 
cubic prepotential.  }
\end{figure*}

However, 
para-hyper-K\"ahler manifolds which are constructed using $c^{3+1}_{3+0}$
can generically not be obtained from $c^{4+0}_{3+0}$. As we will show
in Section 4, the manifolds which are in the image of both 
$c$-maps are precisely those which can be obtained by dimensional reduction
of a five-dimensional (Minkowskian) vector multiplet lagrangian 
with respect to one time-like and one space-like direction. This observation
combines nicely with the results of \cite{CMMS}, where we worked out
and compared 
the dimensional reduction of vector multiplets from $4+1$ to 
$3+1$ dimensions (reduction over space) and to $4+0$ dimensions
(reduction over time). Vector multiplets in $4+1$ dimensions 
contain one real scalar, and the scalar manifolds are so-called
affine very special real manifolds. For such manifolds the metric
is encoded in a real polynomial of degree 3. By dimensional reduction
over space (time) the scalars become complex (para-complex) and
the resulting scalar manifolds are affine special K\"ahler and 
affine special para-K\"ahler, respectively. The dimensional reduction 
defines the so-called $r$-maps, $r^{4+1}_{3+1}$ and $r^{4+1}_{4+0}$ 
between the respective scalar manifolds. If one performs
a reduction from $4+1$ to $3+0$ dimensions, the result is independent
of whether one reduces first over space or time, and the corresponding
para-hyper-K\"ahler manifolds are isometric. This defines 
a map, called the para-$q$-map,
\be
q^{4+1}_{3+0} := c^{3+1}_{3+0} \circ r^{4+1}_{3+1} \cong
c^{4+0}_{3+0} \circ r^{4+1}_{4+0} \;,
\ee
which assigns to each very special real manifold a para-hyper-K\"ahler
manifold. The mutual relations between the various scalar manifolds
are summarized in Figure \ref{Diagram}.

While the physical approach to the $c$-map provides explicit 
expressions for the metric of the hypermultiplet manifold
(and, if we include  the fermions, for the whole lagrangian) its  
geometrical structure is not manifest. If we take, for concreteness,
the dimensional reduction of a Minkowskian lagrangian with $n$
vector multiplets, then we start with $2n$ real scalar fields.
However, the manifold $M$ 
spanned by these scalars is affine special K\"ahler and 
in particular complex. Thus we have a complex structure
$J$, $J^2=-\mathbbm{1}$. After dimensional reduction over time and 
dualization of the vector fields, we have a scalar manifold $N$ of
real dimension $4n$. One can then show that the real scalars can be combined
into $2n$ complex fields, i.e., the complex structure
$J$ of $M$ induces a complex structure $J_1$ on $N$. But this
already requires to identify the proper way of combining 
real fields into complex ones, see Section 3 for the details.
In order
to prove that $N$ is para-hyper-K\"ahler, we have to
work even harder: first, we have to find a para-complex
structure $J_2$, which anticommutes with $J_1$. This gives us
a second para-complex structure $J_3 = J_1 J_2$ for free, and $J_1, J_2, J_3$
satisfy the para-quaternionic algebra. Second, we have to show
that the metric is para-hyper-Hermitian, i.e., $J_2,J_3$ must
be anti-isometries and $J_1$ must be an isometry. Third, 
the metric is even para-hyper-K\"ahler, i.e., the structures
$J_1, J_2, J_3$ must be covariantly constant with respect to the
Levi-Civita connection. The main problem with this approach is
that there is no systematic procedure 
to identify $J_1$ and $J_2$. A strategy used in the 
literature is to show that the metric has the appropriate 
restricted holonomy group \cite{CFG,FS}. Here one uses that a 
Riemannian manifold of dimension
$4n$ is hyper-K\"ahler if and only if its holonomy group is 
contained in $USp(2n)$ (the compact real form of $Sp(\mathbbm{C}^{2n}))$. 
Analogously a Riemannian manifold of 
dimension $4n$ is para-hyper-K\"ahler if and only if its
holonomy group is contained in the real symplectic group 
$Sp(2n,\mathbbm{R}) = Sp(\mathbbm{R}^{2n}) = \rm{Id}_{\mathbbm{R}^2}
\otimes Sp(\mathbbm{R}^{2n}) 
\subset GL(\mathbbm{R}^2 \otimes \mathbbm{R}^{2n}) = 
GL(4n,\mathbbm{R})$.
However, the only efficient way to show that the holonomy group is
contained in $USp(2n)$ and $Sp(2n,\mathbbm{R})$, respectively, is
to show that the structures $J_\alpha$ are parallel. Thus the identification
of these structures is indeed the main problem.

Here lies the great advantage of the geometrical  construction 
which we present in Section 2. The basic result is that 
if $M$ is any special para-K\"ahler manifold (Theorem 2) 
or any special K\"ahler manifold (Theorem 3),\footnote{This
Theorem also applies to the case of indefinite signature
(special pseudo-K\"ahler manifolds). These manifolds are not
admissible target spaces for rigid vector multiplets, but play
an important role when constructing the projective special
K\"ahler manifolds which are the target spaces of locally
supersymmetric vector multiplets \cite{Freed}. This is well
known in the context of the superconformal tensor calculus, 
where the metric of vector multiplet scalars is indefinite before
fixing the gauge for dilatations and $U(1)$ transformations
\cite{deWLauVanP}. 
The indefiniteness of the metric reflects that some of
the fields act as superconformal compensators. We 
expect that pseudo-K\"ahler manifolds will appear, 
when constructing the local version of the para-$c$-map.
This will be addressed in a future publication.}
then its  cotangent bundle $N=T^* M$ is a para-hyper-K\"ahler manifold.
All the geometrical data of $N$, including the structures  
$J_1, J_2, J_3$ can be constructed in terms of the special geometry
data of $M$. The proof of the central Theorems 2 and 3 
requires some preliminaries.
In Section 2.1 we review the relevant facts and definitions 
about para-complex and para-hypercomplex geometry, and about
special (para-)K\"ahler manifolds. Sections 2.2 and 2.3
derive properties of the cotangent bundle $N=T^*M$ of a 
para-complex manifold $M$. In particular, we show that 
$N$ has a canonical para-complex structure $J_N$.
If $M$ is equipped with a 
linear connection $\nabla$, one can decompose the tangent bundle $TN$ of $N$
into a horizontal and vertical part which are isomorphic to
the pullbacks of the 
tangent bundle $TM$ and of the cotangent bundle $T^*M$ of $M$,
respectively. Moreover $\nabla$ can be used to define 
another para-complex structure $J^\nabla$ on $N$. We derive
a sufficent condition for $\nabla$, which implies that $J^\nabla$
is the canonical para-complex structure $J_N$. It turns out 
that this condition is met if $M$ is special para-K\"ahler and
$\nabla$ is its special connection. In Section 2.4 we introduce
a metric on $M$. If $M$ is (almost) para-Hermitian, then one 
can find a metric $g_N$ on $N$ and a second para-complex structure
$J^\omega$ on $N$, such that $N$ is an (almost) para-hyper-Hermitian
manifold. This is the subject of Theorem 1. In Section 2.5 
we take $M$ to be a special para-K\"ahler manifold and prove
Theorem 2, which states that $N$ is para-hyper-K\"ahler. We also
formulate Theorem 3, which claims the same
result if $M$ is special (pseudo-)K\"ahler. The proof is omitted because
it is completely analogous to the one of Theorem 2. 
At this point we have already the complete geometrical picture.
In Sections 2.6 and 2.7 we reformulate the constructions
behind Theorems 2 and 3 in terms of (para-)holomorphic coordinates. 
The action of the structures $J_\alpha$ in the (para-)holomorphic
basis is given in Theorems 4 and 5, 
respectively. We also work out the 
explicit expressions for the metric $g_N$ in terms
of (para-)holomorphic coordinates. In this parametrization
it is easy to compare our results with the Minkowski case \cite{CFG}.

\subsection{Outlook}

Since we only discuss the $c$-map for rigidly supersymmetric theories
in this paper, the natural next step is to consider hypermultiplets
in supergravity. 
It will be interesting to have a fresh look on the 
Minkowskian version of the local $c$-map, because the geometrical
methods used and developed in this paper apply to this case as well. 
We expect that the geometrical structures behind the $c$-map
can be made more transparent. Here
the recent work on superconformal hypermultiplets \cite{deWit:2001dj}
should be helpful. 
Given that not much is known about higher-dimensional
hypermultiplet manifolds in (Minkowskian) supergravity, this is
worthwhile to investigate. Moreover we can treat the Minkowskian and the
Euclidean case in parallel. 

Another extension of the results of this paper is to include the
fermionic degrees of freedom. This is necessary in view of our 
ultimate goal, the construction of supersymmetric instanton solutions.
We also remark that so far we only constructed a particular subset 
of the possible Euclidean hypermultiplet manifolds, namely those in
the image of the para-$c$-map. But by analogy to Minkowskian 
hypermultiplets one expects that any para-hyper-K\"ahler manifold
defines a supersymmetric hypermultiplet lagrangian. 
The corresponding statement for vector multiplets was proven 
in \cite{CMMS}: any special  para-K\"ahler manifold, not just
those in the image of the $r$-map (i.e., those obtainable by
dimensional reduction from five dimensions) defines a Euclidean
vector multiplet lagrangian, and, moreover, these are the most
general admissible scalar manifolds. The idea of the proof was
to rewrite the Euclidean lagrangian and supersymmetry
transformation rules in such a way that they took the same 
form as their Minkowskian counterparts. We expect that this
approach applies to hypermultiplets as well.

\section{Geometrical construction of the para-$c$-maps}

\subsection{Definitions and basic facts}
Let us first recall some definitions and basic facts about para-complex and 
para-\-hyper\-complex geometry, see \cite{CMMS} for more on (special)  
para-K\"ahler manifolds and \cite{ABCV} for (symmetric) para-hyper-K\"ahler 
manifolds.  

\bd An {\sf almost para-complex structure} on a smooth manifold $M$
is an endomorphism field $J \in \G ({\rm End}\,  TM)$, 
such that 
\begin{enumerate} 
\item[(i)] $J \neq {\rm Id}_{TM}$ is an involution, i.e.,\  
$J^2 = {\rm Id}_{TM}$ and 
\item[(ii)] the two eigendistributions  
$T^{\pm}M := {\rm ker} ({\rm Id} \mp J)$ 
of $J$ have the same 
rank.   
\end{enumerate} 
An almost para-complex structure $J$ is called {\sf integrable} if the
distributions $T^{\pm}M$ are both integrable. 
A {\sf para-complex structure} is an integrable almost para-complex structure.
A manifold $M$ endowed with an (almost) para-complex structure $J$ is called
and {\sf (almost) para-complex manifold}. A map $f : (M,J) \ra (M',J')$ 
between (almost) para-complex manifolds is called {\sf para-holomorphic}  
if $df J = J'df$. 
\ed 

Notice that the endomorphism $J_p\in {\rm End}\, T_pM$, $p\in M$, defines 
on the tangent space $T_pM$ the structure of a free module of rank 
$n = \dim T^\pm_pM$ over the ring of para-complex numbers 
\be C := \bR[e] = \{ a+eb\, |\, a, b \in \bR \}\, , \quad e^2=1\, .\ee
The free module $C^k$, $k\in \mathbb{N}$,  is itself an example of a 
para-complex manifold, 
the para-complex structure being the multiplication by $e$. 
In particular, we can speak of {\it para-holomorphic functions}
$f: (M,J) \ra C$.   

Every para-complex manifold $(M,J)$ admits a system of so-called {\it adapted
local coordinates}  $(z^1_+, \ldots, z^n_+, z^1_-,\ldots, z^n_-)$, such that
the $z^i_+$ (respectively, $z^i_-$) are constant on the leaves of the 
integrable distribution $T^-M$ (respectively, $T^+M$).   One can check that
the $C$-valued functions 
\be z^i := \frac{z^i_+ +z^i_-}{2} + e \frac{z^i_+ - z^i_-}{2}\ee 
are para-holomorphic. They form what is called a 
system of {\it para-holomorphic
local coordinates}. 

\bd A pseudo-Riemannian metric $g$ on an almost para-complex manifold
$(M,J)$ is called {\sf para-Hermitian} if $J$ is skew-symmetric
with respect to $g$. An {\sf (almost) para-Hermitian manifold} is an
(almost) para-complex manifold $(M,J)$ endowed with a para-Hermitian
metric $g$. The {\sf fundamental two-form} of an almost para-Hermitian manifold
$(M,J,g)$ is the non-degenerate skew-symmetric bilinear form 
$\o = g(J\cdot , \cdot) $. 
A {\sf para-K\"ahler manifold} is a para-Hermitian manifold $(M,J,g)$ 
for which 
the fundamental two-form $\o$ is closed. In that case, the 
symplectic form  $\o$ is called
the {\sf para-K\"ahler form} of $(M,J,g)$. 
\ed 

Notice that the  skew-symmetry of $J$ with respect to $g$ implies that 
$J$ is an {\it anti-isometry}:
\be J^*g = g(J\cdot , J\cdot ) = -g\, .\ee

As for usual almost Hermitian manifolds, the integrability of $J$ together 
with the para-K\"ahler condition $d\o =0$ on an almost 
para-Hermitian manifold $(M,J,g)$ is equivalent to 
$J$ being parallel for the Levi-Civita connection 
of $(M,g)$. 
  
\bd A {\sf special para-K\"ahler manifold}\footnote{More precisely, 
an {\it affine}  
special para-K\"ahler manifold.} $(M,J,g,\n )$ is a 
para-K\"ahler manifold $(M,J,g)$ endowed with a flat torsion-free
connection $\n$ such that
\begin{enumerate}
\item[(i)] $\n$ is symplectic with respect to the para-K\"ahler form, 
i.e.,\ $\n \o = 0$ and 
\item[(ii)] $\n J$ is a symmetric (1,2)-tensor field, i.e.,\  
$(\n_XJ)Y = (\n_YJ)X$ for all $X,Y$.   
\end{enumerate} 
\ed

It is proven in \cite{CMMS} 
that any simply connected special para-K\"ahler manifold
$(M,J,g,\n )$ admits a canonical realization as a para-holomorphic Lagrangian 
immersion $\phi : M \ra V = C^{2n}$, which induces the special 
geometric structures on $M$. Here $V$ is endowed with the  
standard para-holomorphic symplectic form $\O$ and the standard 
real structure, for which $\bR^{2n} \subset C^{2n}$ is 
the subset of real points. 
 
For any point $p\in M$, there exists a global system of 
linear para-holomorphic coordinates $(z^i,w_i)$ on $V$ which is  
compatible with the structures on $V$ and has the property that 
the image $\phi (U)$ of some neighborhood 
$U\subset M$ of $p$ is defined by a system of equations of the form
\be w_i = F_i := \frac{\partial F}{\partial {z^i}}\, ,\ee
where $F=F(z^1, \ldots, z^n)$ is a (locally defined) 
para-holomorphic function of $n$ variables. $F$ is called 
the para-holomorphic {\it prepotential}.    
The compatibility of $(z^i,w_i)$ with the structures on $V$
means  that the coordinates are 
{\it canonical} for $\O$, i.e.,\ $\O  = \sum dz^i\wedge dw_i$, 
and real-valued on $\bR^{2n} \subset C^{2n}=V$. 

Next we provide the basic definitions of para-hypercomplex geometry. 
\bd An {\sf (almost) para-hypercomplex manifold} is a
smooth manifold $M$ endowed with an  {\sf (almost) para-hypercomplex
structure}, i.e.,\ with three pairwise anticommuting endomorphism fields 
$J_1, J_2, J_3=J_1J_2 \in \G ({\rm End}\,  TM)$,    
such that two of them are (almost) para-complex structures and one of them
is an (almost) complex structure. 

An  {\sf (almost) para-hyper-Hermitian
manifold} is an (almost) para-hypercomplex manifold $(M,J_1,J_2,J_3)$
endowed with a pseudo-Riemannian metric $g$ for which the three
endomorphism fields $J_\a$ are skew-symmetric. The quadruplet  
$(J_1,J_2,J_3,g)$ is called an {\sf (almost) para-hyper-Hermitian
structure}. 

A para-hyper-Hermitian manifold or structure is called
{\sf para-hyper-K\"ahler} if the three fundamental two-forms
$\o_\a = g(J_\a \cdot ,\cdot )$ are closed.    
\ed 
\subsection{The cotangent bundle of a para-complex manifold}
Let $(M,J)$ be a para-complex manifold. Its cotangent bundle 
$N=T^*M$ (as well as its tangent bundle) carries a canonical
para-complex structure $J_N$ such that the canonical projection
$\pi : N \ra M$ is para-holomorphic, i.e.,\
$d \pi J_N =Jd\pi$. Adapted local coordinates for $J_N$
can be constructed as follows. Let $(z^i_\pm )_{i=1,\ldots, n}$ be a system of 
adapted local coordinates for $(M,J)$ defined on an open set $U\subset M$. 
We can consider them as functions on the open set 
$\pi^{-1}(U)\subset N$ via the projection
$\pi : N \ra M$. We define new functions $w_i^\pm$ on 
$\pi^{-1}(U)$ which are linear on the fibers of $\pi$ and
satisfy  
\be w_i^\pm (dz_\pm^j) = \d_i^j\, ,\quad w_i^\pm (dz_\mp^j) = 0\,.\ee 
Then $(z^i_\pm ,w_i^\pm)_{i=1,\ldots, n}$ is a system of 
adapted local coordinates on $(N,J_N)$, i.e.,\
\be \label{JNpartialEqu} J_N \left(\frac{\partial}{\partial z^i_\pm}\right) = 
\pm \frac{\partial}{\partial z^i_\pm}\, ,\quad 
J_N \left( \frac{\partial}{\partial w_i^\pm}\right) = 
\pm \frac{\partial}{\partial w_i^\pm}\, .\ee  
\subsection{The almost para-complex structure on $N$ associated to a
connection on $(M,J)$} 
A linear connection $\n$ on $M$ defines a decomposition
\be \label{decompEqu} 
T_\xi N = {\cal H}^\n_\xi \oplus T^v_\xi N \cong T_pM \oplus T_p^*M\, ,
\quad \xi \in N\, ,\quad p =\pi (\xi)\, ,\ee
where $\pi : N=T^*M \ra M$  is the canonical projection,  the vertical space 
\be T^v_\xi N := {\rm ker}\, d\pi_\xi = T_\xi N_p\cong N_p\ee 
is canonically
identified with the vector space $N_p = T^*_pM$ and the 
horizontal space ${\cal H}^\n_\xi$ is identified with the tangent space to
the basis via the isomorphism 
\be {d\pi_\xi |}_{{\cal H}^\n_\xi } : {\cal H}^\n_\xi \stackrel{\sim}{\ra} 
T_pM\, .
\ee 
Notice that the vertical subbundle $T^vN \subset T N$ is $J_N$-invariant,
since the projection $\pi$ is para-holomorphic,  
whereas the horizontal subbundle  ${\cal H}^\n \subset T N$ is in general
not $J_N$-invariant. This allows us to define a 
second {\it almost} para-complex structure $J^\n$ by
\be \label{JnDefEqu} J_\xi^\n = 
\left(  
\begin{array}{cc} J&0\\
0&J^*\end{array} 
\right)  
\, ,\ee 
with respect to the canonical identification $T_\xi N \cong T_pM 
\oplus T_p^*M$, explained above. Here we used the standard notation
$J^*\a = \a \circ J$, for any $\a\in T_p^*M$. More generally, 
we can define $J^\n$ for any {\it almost} para-complex structure $J$. However, 
in the following we will assume that $J$ is integrable.  
$J^\n$ coincides with the canonical para-complex structure $J_N$ 
only if $J_N{\cal H}^\n = {\cal H}^\n$. In this section we discuss  
necessary and sufficient conditions for the horizontal distribution  
${\cal H}^\n$ to be
$J_N$-invariant.   

\bd A linear connection  $D$ on a para-complex manifold $(M,J)$ is called
a {\sf para-complex connection} if it is torsion-free and satisfies $DJ=0$.
\ed    
\bp \label{nDProp} Let $\n$ be a linear connection on a para-complex manifold
$(M,J)$. Then the horizontal distribution ${\cal H}^\n$ is invariant
under the canonical para-complex structure $J_N$ on $N=T^*M$
if and only if there exists a para-complex connection $D$ on $(M,J)$
such that $A := \n -D$ satisfies 
\be \label{AEqu} A_X \circ J = A_{JX}\, ,\ee
for all $X\in TM$. 
\ep
\pf  
We will use the following standard lemma which relates the horizontal 
distributions with  respect to two connections on the same manifold. 

\bl \label{standardLemma} 
Let $\n$, $D$ be two linear connections on a manifold $M$ and
$A := \n -D\in \O^1({\rm End}\, TM)$ their difference tensor. Then
the corresponding horizontal distributions ${\cal H}^\n$, ${\cal H}^D 
\subset T N$ in the cotangent bundle $N=T^*M$ are related by
\be {\cal H}^\n_\xi = \{ \hat{v} := v + A_{v}^\xi \; |\; v\in
{\cal H}^D_\xi \cong T_pM \}
\, ,\quad \xi \in N\, ,\quad p=\pi (\xi )\, ,\ee
where $A_{v}^\xi := A_{v}^*\xi = \xi \circ A_{v}\in T^*_p M \cong T^v_\xi N$. 
\el
As a first step in the proof of Proposition \ref{nDProp}, 
let us first settle the case when $\n$ is a para-complex connection. 
\bp \label{nProp} 
Let $\n$ be a para-complex connection on a para-complex manifold $(M,J)$.  
Then $J_N{\cal H}^\n = {\cal H}^\n$ and, hence, $J_N=J^\n$.\ep 
\pf Any choice of adapted coordinates $(z^i_\pm )$ on an open set
$U\subset M$ induces a flat para-complex connection $D$ in the
vector bundle  ${T^*M|}_U=T^*U$, such that $Ddz^i_\pm=0$. With respect
to the induced adapted coordinates  $(z^i_\pm ,w_i^\pm)$ on $\pi^{-1}(U)
\subset N$, the horizontal and vertical distributions are simply
\be {\cal H}^D = {\rm span}\{ \frac{\partial}{\partial z^i_\pm}\; |\; 
i=1,\ldots, n\}\quad \mbox{and}\quad 
T^vN =  {\rm span}\{ \frac{\partial}{\partial w_i^\pm}\; |\; 
i=1,\ldots, n\}\, .\ee
In particular, $J_N{\cal H}^D={\cal H}^D$, and $J_N=J^D$, 
cf.\ \re{JNpartialEqu}. 
The tensor $A = \n -D$ satisfies 
\bea A_XY &=& A_YX\label{symmEqu}\\
{[} A_X , J {]} &=& 0\, ,
\eea  
for all tangent vectors $X,Y\in T_pM$, $p\in M$, 
because $\n$ and $D$ are para-complex. The last equation can be 
reformulated as
\be \label{AxiEqu} J^*A_X^\xi = A^{J^*\xi }_X\, ,\ee
for all $\xi \in T^*M$, $X\in T_{\pi (\xi )}M$. 

Now we check that \re{symmEqu} and \re{AxiEqu} imply the $J_N$-invariance
of ${\cal H}^\n$. For any $\hat{v} = v + A_{v}^\xi \in {\cal H}_\xi^\n$,
$v\in {\cal H}^D_\xi \cong T_{\pi (\xi )}M$, see Lemma \ref{standardLemma},  
we calculate
\begin{eqnarray*} J_N\hat{v} &=& J^D(v + A_{v}^\xi ) = Jv + J^*A^\xi_v
\stackrel{\re{AxiEqu}}{=} Jv + A^{J^*\xi }_v \stackrel{\re{symmEqu}}{=}
Jv + A^{J^*\xi }v\\
&\stackrel{\re{AxiEqu}}{=}& 
Jv + A^\xi (Jv) 
\stackrel{\re{symmEqu}}{=}
Jv + A^\xi_{Jv} = \widehat{Jv}\in {\cal H}_\xi^\n\, . 
\end{eqnarray*}  
This proves Proposition \ref{nProp}.  \qed 

Now we prove Proposition \ref{nDProp}. Let $\n$ be a connection
on $(M,J)$ such that $J_N{\cal H}^\n = {\cal H}^\n$. The integrability
of $J$ implies the existence of a para-complex connection $D$ on $(M,J)$. 
In fact, as explained above, any adapted local coordinate system 
on $(M,J)$ defines locally a flat para-complex connection. Pasting these 
locally defined connections by a smooth partition of unity, we obtain 
a globally defined para-complex connection $D$ (which, in general, is not 
flat). By Proposition \ref{nProp}, we have $J_N=J^D$, which allows us to 
compute for $\hat{v} = v + A_{v}^\xi \in {\cal H}_\xi^\n$,
($v\in {\cal H}^D_\xi$, $\xi\in N$):
\[ J_N\hat{v} = J^D(v+A_v^\xi) = Jv + J^*A_v^\xi\, .\] 
Since $J_N{\cal H}^\n = {\cal H}^\n = \{ \hat{v}\; |\; v\in {\cal H}^D\}$, 
we conclude that 
\[  J^*A_v^\xi = A_{Jv}^{\xi}\, .\] 
This proves that $A$ satisfies  \re{AEqu}. 

Conversely, let $D$ be a para-complex connection on $(M,J)$ 
and $\n$ any linear connection on $M$ such that $A=\n -D$ satisfies 
\re{AEqu}. Again, by Proposition \ref{nProp}, we have $J_N=J^D$, and 

\[ J_N\hat{v} = J^D(v+A_v^\xi) = Jv + J^*A_v^\xi \stackrel{\re{AEqu}}{=}
Jv + A_{Jv}^\xi = \widehat{Jv}\, ,\]
for all $v\in {\cal H}^D_\xi$, $\xi\in N$. 
This shows that $J_N{\cal H}^\n = {\cal H}^\n$.  
\qed 

\bp \label{nspecialProp} 
Let $(M,J)$ be a para-complex manifold and $\n$ a torsion-free
connection on $M$ such that $\n J$ is symmetric. Then $J_N{\cal H}^\n 
= {\cal H}^\n$ and, hence, $J_N=J^\n$. 
\ep

\pf The connection 
\be \n^{(J)} := J\circ \n \circ J^{-1} = J\circ \n \circ J = \n +J(\n J)\ee
is torsion-free, since $\n$ is torsion-free and $\n J$ is symmetric. 
Therefore 
\be D := \frac{1}{2}(\n + \n^{(J)}) = \n + \frac{1}{2}J(\n J)\ee
is a torsion-free connection. We claim that $D$ is para-complex.
In fact, for all $X\in TM$, we have 
\begin{eqnarray*} D_XJ &=&
 \n_XJ + \frac{1}{2}[J(\n_X J),J]\\  
&=&  \n_XJ + \frac{1}{2}(J(\n_X J)J -\n_X J)\\ 
&=& \n_XJ -\n_X J = 0\, .
\end{eqnarray*} 
Here we used that $J^2= {\rm Id}$ and, thus, $0 = \n_X J^2 = (\n_X J)J + 
J\n_XJ$. 
Now, in view of Proposition \ref{nDProp}, it suffices to show that
\be A=\n -D = -\frac{1}{2}J(\n J)\ee
 satisfies \re{AEqu}, which is equivalent to 
\be (\n_{X}J)\circ J = \n_{JX}J\, , \quad \mbox{for all}\quad X\in TM\, .\ee
This follows from the symmetry of $\n J$:
\[ (\n_{X}J)\circ J = -J \circ (\n_X J) = -J(\n J)X = (\n J)JX 
= \n_{JX}J\, .\]
\qed 
\subsection{The almost para-hyper-Hermitian structure on $N$ associated to a
connection and a para-Hermitian metric on $(M,J)$} 
Let $(M,J,g)$ be an almost para-Hermitian manifold and $\n$ a linear
connection on $M$. Then we have the almost para-complex structure
$J^\n$ on $N=T^*M$, see \re{JnDefEqu}. 
Using the fundamental two-form $\o = g(J \cdot, \cdot )$, 
we define  a second almost para-complex structure $J^\o$ on $N$ by 
\be J_\xi^\o = 
\left(  
\begin{array}{cc} 0&\o^{-1}\\
\o & 0\end{array} 
\right)  
\, ,\ee 
with respect to the canonical identification $T_\xi N \cong T_pM 
\oplus T_p^*M$ \re{decompEqu}. We define also a pseudo-Riemannian metric
$g_N$ on $N$ by 
\be \label{c4030metricEqu} g_N= 
\left(  
\begin{array}{cc} g&0\\
0& g^{-1}\end{array} 
\right)  
\, .\ee 

\bt \label{almostThm} 
Let $(M,J,g)$ be an almost para-Hermitian manifold endowed with 
a linear connection $\n$. Then $(N, J_1:=J^\n, J_2:=J^\o, J_3:=J_1J_2, g_N)$ 
is an almost para-hyper-Hermitian manifold.    
\et 

\pf The skew-symmetry of $J$ with respect to $\o$ can be written as
\be J^*\circ \o = -\o \circ J\, ,\ee
if $\o$ is considered as a map 
\bea \o &:& TM \ra T^*M\\ 
&& v\mapsto \o(v,\cdot)\nonumber\, .
\eea
As a consequence, we obtain 
$\o^{-1}\circ J^* = - J\circ \o^{-1}$ and, hence, the identity 
$J_1J_2 = -J_2J_1$. This proves that $(J_1, J_2, J_3)$ is an
almost para-hypercomplex structure. It is clear that $J_1$ is
skew-symmetric with respect to $g_N$, since $J$ is $g$-skew-symmetric.
It remains to check that $J_2$ is $g_N$-skew-symmetric. We have 
to check that the bilinear form 
\be \label{o2Equ} g_N\circ J_2 = \left(  
\begin{array}{cc} g&0\\
0& g^{-1}\end{array} 
\right) \left(  
\begin{array}{cc} 0&\o^{-1}\\
\o &0 \end{array} 
\right)   = 
\left(  
\begin{array}{cc} 0&g\circ \o^{-1}\\
g^{-1}\circ \o & 0\end{array} 
\right)
= \left(  
\begin{array}{cc} 0&g\circ \o^{-1}\\
J& 0\end{array} 
\right) 
\ee
is skew-symmetric. In other words, we have to check that
\be \label{goEqu}(g\circ \o^{-1})^* = -J\, .\ee 
Notice that the symmetry of $g$ and 
skew-symmetry of $\o$, can be expressed as 
\be g^* = g \quad \mbox{and}\quad  \o^* = -\o : TM = (T^*M)^*\ra T^*M\, .\ee
This yields the desired identity 
\[  (g\circ \o^{-1})^* = (\o^{-1})^*\circ g^* = -\o^{-1}\circ g = 
-(g^{-1}\circ \o )^{-1} = -J^{-1} = -J\, .\]    
\qed 
\subsection{The para-hyper-K\"ahler structure on the cotangent bundle
of a special para-K\"ahler manifold}   
\bt \label{c40/30Thm} Let $(M,J,g,\n )$ be a special para-K\"ahler manifold with para-K\"ahler
form $\o$ and $(J_1, J_2, J_3, g_N)$ 
the almost para-hyper-Hermitian structure on $N=T^*M$ constructed in 
Theorem \ref{almostThm}. Then  $(N, J_1, J_2, J_3, g_N)$  is a 
para-hyper-K\"ahler manifold. 
\et    

\pf We have to show that the two almost para-complex structures $J_1$, $J_2$
and the almost complex structure $J_3$ are integrable and that the
two-forms $\o_\a := g_N(J_\a \cdot ,\cdot )$, $\a = 1,2,3$, are closed. 
By Proposition \ref{nspecialProp}, $J_1=J^\n$  coincides with the
canonical (integrable) para-complex structure $J_N$. Since the
symplectic form $\o$ is $\n$-parallel, the almost para-complex 
structure $J_2=J^\o$ is represented by a constant matrix with respect
to any local $\n$-affine coordinate system. This proves the integrability
of $J_1$ and $J_2$. Let us now check that the corresponding
fundamental two-forms are closed. With respect to the canonical 
identification $TN={\cal H}^\n \oplus T^v N\cong TM \oplus T^*M$, we 
have 
\be \label{o1Equ} \o_1 = g_NJ_1 = \left(  
\begin{array}{cc} g&0\\
0& g^{-1}\end{array} 
\right)  \left(  
\begin{array}{cc} J&0\\
0&J^*\end{array} 
\right)  = 
\left(  
\begin{array}{cc} g\circ J&0\\
0& g^{-1}\circ J^*\end{array} 
\right) =  
\left(  
\begin{array}{cc} \o &0\\
0& -\o^{-1}\end{array} 
\right) \, ,
\ee 
where we have used the identity \re{goEqu}, which implies $g^{-1}\circ J^*
=g^{-1}\circ (-g\circ \o^{-1})=-\o^{-1}$. The equation \re{o1Equ} shows
that $\o_1$ has constant coefficients with respect 
to any local $\n$-affine coordinate system and is, therefore, closed.
According to \re{o2Equ}, we have that 
\be \label{o2'Equ} \o_2 = \left(  
\begin{array}{cc} 0&-J^*\\
J& 0\end{array} 
\right) \, .\ee  
To see that $\o_2$ is closed, let us choose $\n$-affine local coordinates  
$(q^i)_{i=1,\ldots,2n}$ on $M$ and express $J$ in these coordinates:
\be \label{JEqu} J\left(\frac{\partial}{\partial q^i}\right)  
= \sum J_i^j\frac{\partial}{\partial q^j}\, . 
\ee 
The expression for $J^*$ in the dual co-frame $(dq^i)$ is 
the transposed matrix of $(J^i_j)$:
\be \label{dualEqu} J^*dq^i =   
\sum J_j^idq^j\, . 
\ee 
The canonical 
isomorphism $T_\xi^v N \cong T_p^*M$, $\xi\in N$, $p = \pi (\xi )$, 
maps 
\be {\left. \frac{\partial}{\partial p_i}\right|}_\xi\mapsto {dq^i|}_p\, ,
\ee 
where $(q^i=q_N^i=\pi^*q^i_M,p_i)$ is the canonical system of local 
coordinates of
$N=T^*M$ associated to the local coordinates $(q^i=q_M^i)$.
Together with
the equations \re{o2'Equ}, \re{JEqu} and \re{dualEqu} this
shows that 
\be \o_2 = \sum J^i_jdq_N^j \wedge dp_i = \sum  
\pi^* (J^*dq_M^i)\wedge dp_i\, .\ee
{}From $\n dq^i=0$ and the symmetry of $\n J$, we obtain that
the one-forms $J^*dq^i$ on $M$ are closed:
\be dJ^*dq^i = {\rm alt} \n (dq^i \circ J) = (dq^i) \circ {\rm alt}\n J = 0\, .
\ee 
Here we used the fact that the exterior derivative is the 
alternation (anti-symmetrization, up to a factor depending on the 
conventional identification between totally skew-symmetric tensors and 
exterior forms) 
of the covariant derivative 
for any torsion-free connection.  
Since the pull back of any closed form is closed, we obtain that the 
one-forms $\pi^* (J^*dq^i)$ on $N$ are closed. This proves that $\o_2$ 
is closed. So we have proven that $(N,J_1,g_N)$ and $(N,J_2,g_N)$ are
para-K\"ahler manifolds. In particular, $J_1$ and $J_2$ are
parallel with respect to the Levi-Civita connection of $(N,g_N)$. 
Therefore, $J_3=J_1J_2$ is also parallel, and $(N,J_3,g_N)$ is an 
indefinite K\"ahler manifold. 
\qed   

Similarly, we can prove the following theorem. 
\bt \label{c31/30Thm} 
Let $(M,J,g,\n )$ be a special (pseudo-)K\"ahler manifold with K\"ahler
form $\o$. Then the cotangent bundle $N=T^*M$ carries a 
para-hyper-K\"ahler structure $(J_1, J_2, J_3, g_N)$, associated to the 
special K\"ahler structure on $M$. With respect to the canonical 
identification \re{decompEqu}, the complex structure $J_1$ is given by 
\be J_1 := J^\n = J_N = \left(  
\begin{array}{cc} J&0\\
0&J^*\end{array} 
\right)\, ,\ee
the two para-complex structures $J_2$, $J_3$ are  
\be J_2 := J^\o = \left(  
\begin{array}{cc} 0&\o^{-1}\\
\o & 0\end{array} 
\right)\, ,\quad 
J_3 := J_1J_2\ee
and the para-hyper-K\"ahler metric is
\be g_N  := \left(  
\begin{array}{cc} g&0\\
0& -g^{-1}\end{array} 
\right) \, .
\label{HKM}
\ee    
\et    

Summarizing, we have defined two maps 
\bea c=c^{4+0}_{3+0}&:&\{ \mbox{special para-K\"ahler manifolds}\} \ra 
\{ \mbox{para-hyper-K\"ahler manifolds}\}\\
c=c^{3+1}_{3+0}&:&\{ \mbox{special K\"ahler manifolds}\} \ra 
\{ \mbox{para-hyper-K\"ahler manifolds}\}\\
&& (M,J,g,\n ) \mapsto c(M,J,g,\n ) := (N,J_1, J_2, J_3, g_N)\, ,\eea
which we call the (affine) {\sf para-$c$-maps}. We can compose them
with the two $r$-maps 
\bea r^{4+1}_{3+1}&:&\{ \mbox{affine very special real manifolds}\} \ra 
\{ \mbox{special K\"ahler manifolds}\}\\
r^{4+1}_{4+0}&:&\{ \mbox{affine very special real manifolds}\} \ra 
\{ \mbox{special para-K\"ahler manifolds}\}\eea
(see \cite{CMMS} and references therein), obtaining two maps
\[ q^{4+1}_{3+0} := c^{3+1}_{3+0}\circ r^{4+1}_{3+1}\quad \mbox{and} \quad 
c^{4+0}_{3+0}\circ r^{4+1}_{4+0} :\] 
\be \{ \mbox{affine very special real manifolds}\} \ra 
\{ \mbox{para-hyper-K\"ahler manifolds}\}\, .\ee 
We call $q^{4+1}_{3+0}$ {\it the}  {\sf para-$q$-map}. This is 
justified by the following proposition, which follows from the 
discussion in section \ref{5to3}. 

\bp For any affine very special real manifold $L$, the  
para-hyper-K\"ahler manifolds $(c^{3+1}_{3+0}\circ r^{4+1}_{3+1}) (L)$ and 
$(c^{4+0}_{3+0}\circ r^{4+1}_{4+0})(L)$ are canonically isometric.
\ep  

\noindent 
{\bf Remark 1:} \label{rem1} 
Let  $(M,J,g,\nabla )$ be a special (para-)K\"ahler manifold. 
Then 
\be (M,J,cg,\nabla^{(a,b)})\, ,\quad   \nabla^{(a,b)} := 
(a{\rm Id}+bJ)\circ \nabla \circ  
(a{\rm Id}+bJ)^{-1}\, , 
\ee 
is again a 
special (para-)K\"ahler manifold, for all $a,b,c\in \bR$ such that
\be (a{\rm Id}+bJ)(a{\rm Id}-bJ) = \pm {\rm Id}
\quad \mbox{and}\quad c\neq 0\, .\ee   
As a consequence, applying the para-$c$-map to 
the (non-connected) family $(M,J,cg,\nabla^{(a,b)})$  provides 
a family of para-hyper-K\"ahler structures on $N=T^*M$, which 
depends on two parameters. In addition, we also have   
the trivial freedom of multiplying
the metric $g_N$ by a non-zero constant.   
 
Comparing the formulae  \re{c4030metricEqu} and \re{HKM} for the 
para-hyper-K\"ahler metrics $g_N$ obtained by the para-$c$-maps
$c^{4+0}_{3+0}$ and $c^{3+1}_{3+0}$, respectively, one may wonder
about the different sign in front of the inverse metric. As we shall 
explain now, the plus sign in  front of $g^{-1}$ in \re{c4030metricEqu}
can be converted into minus by a $J_1$-holomorphic diffeomorphism
$\psi$ of $N$.     
In fact, the map $J^* : T^*M \ra T^*M$ defines a diffeomorphism
$\psi$ of $N=T^*M$. Its differential $d\psi$ 
preserves the vertical distribution $T^vN$ and maps the 
horizontal distribution ${\cal H}^\n$ with respect to the 
connection $\n$ to the  horizontal distribution 
${\cal H}^{\n '}$ with respect to the 
connection $\n'=J\circ \n \circ J^{-1}$. 
The differential $d\psi_\xi : T_\xi N \ra T_\xi N$ at the point 
$\xi\in N$ is given by 
\be \label{dpsiEqu} d\psi_\xi = \left(  
\begin{array}{cc} \id &0\\
0&J^*\end{array} 
\right)\, ,\ee
where the domain and target are identified as \re{decompEqu} and 
\be \label{decomp'Equ} T_\xi N = {\cal H}^{\n'}_\xi \oplus T^v_\xi N 
\cong T_pM \oplus T_p^*M\, , \ee
respectively. Using this, one can easily check that
the diffeomorphism $\psi$ transforms the
para-hyper-K\"ahler structure  $(g_N,J_1,J_2,J_3)$
defined by the para-$c$-map $c^{3+1}_{3+0}$, with respect to 
\re{decompEqu},
 trivially to $(g_N,J_1,J_3,-J_2)$, with respect to \re{decomp'Equ},
 and maps the para-hyper-K\"ahler structure  $(g_N = {\rm 
diag}(g,g^{-1}),J_1,J_2,J_3)$
obtained by the para-c-map $c^{4+0}_{3+0}$ to a new para-hyper-K\"ahler 
structure 
$(g_N' = {\rm diag}(g,-g^{-1}),J_1'=J_1,J_2',J_3')$, where
\be J_2' = -\left( 
\begin{array}{cc} 0 &g^{-1}\\
g&0\end{array}
\right) \quad \mbox{and}\quad J_3' = J_1'J_2' = \left( 
\begin{array}{cc} 0 &-\o^{-1}\\
\o &0\end{array}
\right)\, ,\ee
with respect to  \re{decomp'Equ}.

\label{cmapsection} 
\subsection{The Euclidean para-$c$-map in the para-holomorphic 
para\-me\-tri\-za\-tion} 
Later on we shall compare the para-$c$-map $c^{4+0}_{3+0}$ introduced 
in the previous section to the sigma models which occur in 
the dimensional 
reduction of the Euclidean vector multiplet lagrangian from four to three 
dimensions, followed by dualization into a hypermultiplet lagrangian. 
We shall find that they coincide up to a simple 
redefinition using the freedom discussed in \ref{cmapsection} Remark 1. 
For this it is useful to express the geometric data in terms of  
canonical para-holomorphic coordinates, which correspond to the 
para-complex scalar fields obtained from the dimensional 
reduction. This is accomplished in this section. 
The analogous results in the Minkowskian case 
are given in the next section. 

Let $(M,J,g,\n )$ be a special para-K\"ahler manifold and 
$(J_1, J_2, J_3, g_N)$
the para-hyper-K\"ahler structure on $N=T^*M$ constructed in Theorem 
\ref{c40/30Thm}. The geometric structure of $M$ is locally completely specified
by the para-holomorphic prepotential $F = F(z)$ with respect to some
system of special para-holomorphic coordinates $z = (z^i)$. 
 Let $(z^i,w_i)$ be
the canonical $J_1$-para-holomorphic local coordinates of 
\be \label{10Equ} N = T^*M \cong 
\wedge^{1,0}T^*M \subset T^*M \otimes C = \wedge^{1,0}T^*M \oplus 
\wedge^{0,1}T^*M\ee 
associated to the
special coordinates $(z^i)$ on $M$. 
\bp \label{J2Prop} 
The expression for the para-complex structure $J_2$ on $N$ in  the 
canonical para-holomorphic coordinates $(z^i,w_i)$ is the 
following:
\be \label{ThmEqu} J_2^*dz^i = -2e\sum N^{ij}(d\Bar{w}_j -
e\Bar{F}_{jkl}N^{km}(w_m-\Bar{w}_m)d\Bar{z}^l) \, ,\ee
where $(N^{ij})$ is the inverse of the matrix $(N_{ij})$, 
with matrix elements $N_{ij} = e(F_{ij}-\Bar{F}_{ij})$.     
\ep    

\pf Let us first express the para-complex structure $J_2$ in 
canonical coordinates $(q^a,p_a)$ on $N=T^*M$ associated
to a system of $\n$-affine local coordinates $(q^a)$ on $M$. 
We choose 
\be \label{qpEqu} (q^a,p_a) = (x^i,y_i,\hat{x}_i,\hat{y}^i)\, ,\ee 
where 
\be z^i = x^i + eu^i\quad \mbox{and}\quad 
F_i = y_i + ev_i\, .\ee 
Then the para-K\"ahler form
$\o = g(J\cdot ,\cdot )$ of $M$ takes the form 
\be \o = 2\sum dx^i\wedge dy_i\ee 
and, hence,  
\be \label{J2dxdyEqu} J_2^*dx^i = \frac{1}{2}d\hat{y}^i\, ,\quad J_2^*dy_i 
= -\frac{1}{2}d\hat{x}_i \, .
\ee 
This allows us to compute the left-hand side of \re{ThmEqu}: 
\bea \nonumber J_2^*dz^i &=& J_2^*(dx^i + edu^i) = \frac{1}{2}d\hat{y}^i
+ \frac{e}{2}\left(\frac{\partial u^i}{\partial x^j}\,d\hat{y}^j 
-\frac{\partial u^i}{\partial y_j}\,d\hat{x}_j\right)\\ 
&=& 
-\frac{e}{2}\frac{\partial u^i}{\partial y_j}d\hat{x}_j 
+ \frac{1}{2} \left(\delta^i_j + e\frac{\partial u^i}{\partial x^j}\right)d\hat{y}^j \, .
\label{J2dzEqu}\eea 
Notice that, in order to lighten the  calculations, 
we are using Einstein's 
summation convention.  
The imaginary part of the equation $ddF=0$ yields the useful identity
\be \label{u/x=-v/yEqu} \frac{\partial u^i}{\partial x^j} 
= -\frac{\partial v_j}{\partial y_i}\, ,  \ee
which shows that 
\be \label{deltadudxEqu} \delta^i_j + e\frac{\partial u^i}{\partial x^j} = 
 \delta^i_j -e\frac{\partial v_j}{\partial y_i}
= \frac{\partial}{\partial y_i} \Bar{F}_j = \Bar{F}_{jk}
\frac{\partial \Bar{z}^k}{\partial y_i} = 
-e\Bar{F}_{jk}\frac{\partial u^k}{\partial y_i}\, . 
\ee   
Decomposing the last equation in real and imaginary parts
yields 
\be \label{RNEqu} R_{jk}\frac{\partial u^k}{\partial y_i} = -
2\frac{\partial u^i}{\partial x^j} \quad \mbox{and} \quad 
N_{jk}\frac{\partial u^k}{\partial y_i} = 2\delta_j^i\, ,
\ee 
where 
\be R_{ij} := F_{ij} + \Bar{F}_{ij}\, .\ee 
In particular, we have 
\be \label{duoverdyEqu}\frac{\partial u^i}{\partial y_j} = 
2N^{ij} \, .
\ee 
Using \re{J2dzEqu}, \re{deltadudxEqu} and \re{duoverdyEqu} we can rewrite 
\bea \label{final J2dzEqu} \nonumber J_2^*dz^i &=& -eN^{ij}d\hat{x}_j 
-\frac{e}{2}\Bar{F}_{jk}\frac{\partial u^k}{\partial y_i}d\hat{y}^j\\ 
&=& -eN^{ij}(d\hat{x}_j + \Bar{F}_{jk}d\hat{y}^k) \, .
\eea  

In order to check that this coincides with the
right-hand side of \re{ThmEqu}, let us first rewrite 
$w_i$ as a function of the real canonical coordinates  
$(q^a,p_a) = (x^i,y_i,\hat{x}_i,\hat{y}^i)$. The real canonical 
coordinates $(q^a,p_a)$  
are related to the para-holomorphic canonical coordinates $(z^i,w_i)$
via the identification 
\be T^*M \stackrel{\sim}{\ra} 
\wedge^{1,0}T^*M\, ,\quad  
\a = p_adq^a \mapsto \frac{1}{2}(\a +eJ^*\a ) = w_idz^i\, .\ee    
We have 
\be \a = p_adq^a = \hat{x}_idx^i + \hat{y}^idy_i = 
\frac{1}{2}\hat{x}_idz^i + 
\frac{1}{2}\hat{y}^iF_{ij}dz^j + 
\mbox{c.c.} \, .\ee  
Similarly, using the equations 
\be J^*dx^i=du^i\, ,\quad J^*dy_i=dv_i\, ,\ee 
one obtains 
\be eJ^*\a =  \frac{1}{2}\hat{x}_idz^i + \frac{1}{2}\hat{y}^iF_{ij}dz^j -  
\mbox{c.c.} \, .\ee 
Thus 
\be \label{wIdzIEqu} w_idz^i = \frac{1}{2}(\a +eJ^*\a ) = 
\frac{1}{2}(\hat{x}_i + \hat{y}^jF_{ij})dz^i\, .\ee
This shows that 
\be w_i = w_i(x,y,\hat{x},\hat{y}) = \frac{1}{2}( \hat{x}_i + \hat{y}^j F_{ij})
\, ,\ee
where $F_{ij} = F_{ij}(z)$ and $z^i=z^i(x,y)$. In particular, 
\be e(w_i -\Bar{w}_i) = \frac{1}{2}\hat{y}^jN_{ji}\ee 
and, hence, 
\be \label{E_ipK}
dw_i -
eF_{ijk}N^{jl}(w_l-\Bar{w}_l)dz^k =
dw_i -\frac{1}{2}\hat{y}^jF_{ijk}dz^k 
= \frac{1}{2}(d\hat{x}_i + F_{ij}d\hat{y}^j)\, .\ee 
Therefore, the right-hand side of \re{ThmEqu} is given by 
\be -2eN^{ij}(d\Bar{w}_j -
e\Bar{F}_{jkl}N^{km}(w_m-\Bar{w}_m)d\Bar{z}^l)  = -eN^{ij}(d\hat{x}_j + 
\Bar{F}_{jk}d\hat{y}^k)\, .\ee 
Comparing this with \re{final J2dzEqu} yields \re{ThmEqu}. \qed 

In order to compute the para-hyper-K\"ahler metric of $N$ in the 
para-holomorphic coordinates $(z^i,w_i)$, it is useful to introduce 
a para-unitary co-frame  
\be \label{ei} e^i = \sum e^i_Idz^I\, ,\ee 
with respect to 
the sesquilinear para-Hermitian metric\footnote{The minus sign is due to
the conventions of Ch.\ 2 of \cite{CMMS}, see equation (2.2) there.} 
\be 
h := g+e\o = - \sum N_{IJ}dz^Id\Bar{z}^J \ee 
on $(M,J)$, i.e.,\   
\be  \label{NIJ} N_{IJ} = -\sum e^i_I{\Bar{e}^i}_J\, .\ee
{}From now on,
we shall distinguish the holonomic from the para-unitary co-frame
by capital and lower case indices, respectively. 
We also put
\be \label{Ei} E_i := 
2\sum e_i^I(dw_I-eF_{IJK}N^{JL}(w_L-\Bar{w}_L)dz^K)\, ,\quad 
(e_i^I) = (e_I^i)^{-1}\, .\ee 
Notice that $\sum e^i\wedge E_i = 2\sum dz^I\wedge dw_I$. 

Now we express the full para-hyper-K\"ahler structure 
on $N$ in the co-frame $(e^i,E_i)$ of $\wedge^{1,0}_{J_1}T^*N$. 
\bt Let $(M,J,g,\n )$ be a special para-K\"ahler manifold and 
$(J_1, J_2, J_3, g_N)$ the para-hyper-K\"ahler structure on $N=T^*M$ 
constructed in Theorem 
\ref{c40/30Thm}. In the co-frame $(e^i,E_i)$, the 
para-hyper-K\"ahler structure has the following expression:    
\begin{enumerate}
\item[(i)]  The para-hypercomplex structure 
$(J_1,J_2,J_3=J_1J_2)$ is given by 
\bea J_1^*e^i &=& ee^i\, ,\quad J_1^*E_i = eE_i\\
     J_2^*e^i &=& e\Bar{E}_i \, ,\quad 
J_2^*E_i = -e\Bar{e}^i\, .\eea  
\item[(ii)] The $J_1$-sesquilinear 
para-Hermitian metric $h_N := g_N +e\o_1$, $\o_1=g_N(J_1\cdot, \cdot)$,  
is given by 
\be h_N = \sum (e^i\Bar{e}^i + E_i\Bar{E}_i)\, .\ee
\end{enumerate} 
\et 

\pf (i) follows from Proposition \ref{J2Prop}.\\ 
(ii) $h_N$ is the unique $J_1$-sesquilinear form such that 
${\rm Re}\, h_N = g_N$. Therefore it is sufficient to check
that 
\be g_N = {\rm Re}  \sum (e^i\Bar{e}^i + E_i\Bar{E}_i)\, .
\label{dsmReal0}
\ee
In order to check this, we calculate the right-hand side in the
real coordinates (\ref{qpEqu}). 
The first term is
\bea
\label{dsmReal}
g & = & {\rm Re} \sum e^i \Bar{e}^i \\  &=& {\rm Re}\,  
(- N_{IJ} dz^I d \overLine{z}^J) \nonumber \\ &=&  
- \left( N_{IJ} - \ft{\der u^K}{\der x^I} N_{KL} \ft{\der u^L}{\der x^J}
\right)
dx^I dx^J  
+ 2 \left( \ft{\der u^K}{\der x^I} N_{KL} \ft{\der u^L}{\der y^J}
\right) dx^I dy_J  
+ \left( \ft{\der u^K}{\der y^I} N_{KL} \ft{\der u^L}{\der y^J}
\right) dy_I dy_J \;. \nonumber
\eea
The second term is
\bea 
\label{dsmReal2}
 & & 
 {\rm Re} \sum E_i \Bar{E}_i  \\ &=&
-4 N^{IJ} ( dw_I - e F_{IKL} N^{LM}(w_M - \overline{w}_M) dz^L)
( d \overline{w}_J - e \overline{F}_{JNP} N^{NQ}(w_Q - \overline{w}_Q) 
d \overline{z}^P)
\nonumber \\  
&=&   
-N^{IJ} d \hat{x}_I d \hat{x}^J 
- N^{IK} R_{KJ} d \hat{x}_I d \hat{y}^J 
+ \ft14 ( N_{IJ} - R_{IK} N^{KL} R_{LJ} ) d \hat{y}^I d 
\hat{y}^J \;, \nonumber
\eea
where we used \refeq{E_ipK}.
Observe that \refeq{dsmReal} and  \refeq{dsmReal2}
do not contain terms which mix 
$(dx^I, dy_J)$ with $(d\hat{x}_I, d\hat{y}^J)$. For the following
manipulations it is convenient to use matrix notation.
Defining
\be\label{PQdef}
P = (P^I_J) = (\ft{\der u^I}{\der x^J}) \;,\;\;\;
Q = (Q^{IJ}) = (\ft{\der u^I}{\der y_J}) \;.
\ee
we can write 
\be
{\rm Re}  \sum (e^i\Bar{e}^i + E_i\Bar{E}_i) = \left( \begin{array}{cc}
g & 0 \\ 0 \;\; & g' \\
\end{array} \right) \;, 
\ee
where 
\be
g = \left( \begin{array}{cc}
- N + P^T N P \; \; &  P^T N Q \\
Q^T N P &  Q^T N Q \\
\end{array} \right)
\ee
and
\be
g' = -\left( \begin{array}{cc}
 N^{-1} &  \ft12 N^{-1} R \\
 \ft12 R N^{-1} \;\;& - \ft14 \left( N - R N^{-1} R \right) \\
\end{array} \right) \;.
\ee
Now we use the two identities (\ref{RNEqu})
\be\label{RNMEqu}
R Q = - 2 P^T \;,\;\;\; N Q =  2 \; \mathbbm{1} \;,
\ee
in order to rewrite $g$:
\be
g = \left( \begin{array}{cc}
- N + R N^{-1} R  \;\;& - 2 R N^{-1}  \\
- 2 N^{-1} R &  4 N^{-1} 
\end{array} \right)
\ee
Then it is easy to see that $g \, g' = +\mathbbm{1}$,
so that the right-hand side of \refeq{dsmReal0} takes the form
\be
\left( \begin{array}{cc}
g & 0 \\ 0 \;\; & g^{-1} \\
\end{array} \right) = g_N\, .
\ee
\qed \\

\subsection{The Minkowskian para-$c$-map in the holomorphic 
para\-me\-tri\-za\-tion}
For completeness and future use, we now extend the results of the 
previous section to the case of the second para-$c$-map $c^{3+1}_{3+0}$.
We shall see later that this corresponds  to the dimensional
reduction of the four-dimensional Minkowskian vector multiplet lagrangian 
along a time-like direction to three Euclidean dimensions,
followed by the dualization into a hypermultiplet lagrangian. 

Let $(M,J,g,\n )$ be a special K\"ahler manifold and 
$(J_1, J_2, J_3, g_N)$
the para-hyper-K\"ahler structure on $N=T^*M$ constructed in Theorem 
\ref{c31/30Thm}. We denote by $F$ the holomorphic prepotential 
with respect to some local 
system of special holomorphic coordinates $(z^i)$
on $M$ and by $(z^i,w_i)$ the corresponding 
canonical $J_1$-holomorphic local coordinates of 
$N = T^*M \cong 
\wedge^{1,0}T^*M \subset T^*M \otimes \bC$. 
\bp \label{J2Prop'} 
The expression for the para-complex structure $J_2$ on $N$ in  the 
canonical holomorphic coordinates $(z^i,w_i)$ is the 
following:
\be \label{ThmEqu'} J_2^*dz^i = -2\sqrt{-1}\sum N^{ij}(d\Bar{w}_j -
\sqrt{-1}\Bar{F}_{jkl}N^{km}(w_m-\Bar{w}_m)d\Bar{z}^l) \, ,\ee
where $(N^{ij})$ is the inverse of the matrix $(N_{ij})$, 
with matrix elements $N_{ij} = \sqrt{-1}(F_{ij}-\Bar{F}_{ij})$.     
\ep    

\pf Let us first express the para-complex structure $J_2$ in the 
canonical coordinates $(x^i,y_i,\hat{x}_i,\hat{y}^i)$ associated
to the $\n$-affine local coordinates $(x^i,y_i)$  on $M$,  
where now 
\be z^i = x^i + \sqrt{-1}u^i\quad \mbox{and}\quad 
F_i = y_i +  \sqrt{-1}v_i\, .\ee 
If the K\"ahler form of $M$ is defined as
$\o = g(\cdot , J\cdot )$, it takes the same form 
\be \o = 2\sum dx^i\wedge dy_i\ee 
and, hence,  \re{J2dxdyEqu} still holds. 
This allows us to compute the left-hand side of \re{ThmEqu'}: 
\be J_2^*dz^i = 
-\frac{\sqrt{-1}}{2}\frac{\partial u^i}{\partial y_j}d\hat{x}_j 
+ \frac{1}{2} (\delta^i_j + 
\sqrt{-1}\frac{\partial u^i}{\partial x^j})d\hat{y}^j \, .
\label{J2dzEqu'}\ee 
Using the identity \re{u/x=-v/yEqu}, which holds for special K\"ahler 
manifolds, as well
as for special para-K\"ahler manifolds,  we obtain now that 
\be \label{deltadudxEqu'} \delta^i_j + 
\sqrt{-1}\frac{\partial u^i}{\partial x^j} = 
-\sqrt{-1}\Bar{F}_{jk}\frac{\partial u^k}{\partial y_i}\, . 
\ee   
Decomposing the last equation in real and imaginary parts
yields again \re{RNEqu} and, in particular,  
\be \label{duoverdyEqu'}\frac{\partial u^i}{\partial y_j} = 
2N^{ij} \, .
\ee 
Using \re{J2dzEqu'}, \re{deltadudxEqu'} and \re{duoverdyEqu'} we can rewrite 
\be \label{final J2dzEqu'} J_2^*dz^i = 
-\sqrt{-1}N^{ij}(d\hat{x}_j + \Bar{F}_{jk}d\hat{y}^k) \, .
\ee  

In order to check that this coincides with the
right-hand side of \re{ThmEqu'}, let us again rewrite 
$w_i$ as a function of the coordinates  
$(q^a,p_a) = (x^i,y_i,\hat{x}_i,\hat{y}^i)$. The real coordinates $(q^a,p_a)$  
are now related to the holomorphic coordinates $(z^i,w_i)$
via the identification 
\be T^*M \stackrel{\sim}{\ra} 
\wedge^{1,0}T^*M\, ,\quad  
\a = p_adq^a \mapsto \frac{1}{2}(\a -\sqrt{-1}J^*\a ) = w_idz^i\, .\ee    
Using the equations 
\be J^*dx^i=-du^i\, ,\quad J^*dy_i=-dv_i\, ,\ee
we now obtain    
\bea \a &=& 
\frac{1}{2}\hat{x}_idz^i + 
\frac{1}{2}\hat{y}^iF_{ij}dz^j + 
\mbox{c.c.}\\ 
 \sqrt{-1}J^*\a 
&=&  -(\frac{1}{2}\hat{x}_idz^i + \frac{1}{2}\hat{y}^iF_{ij}dz^j -  
\mbox{c.c.}) \eea  
and, thus,  as before,  
\be w_idz^i = \frac{1}{2}(\a -\sqrt{-1}J^*\a ) = 
\frac{1}{2}(\hat{x}_i + \hat{y}^jF_{ij})dz^i\, .\ee
This shows, in particular, that 
\be \sqrt{-1}(w_i -\Bar{w}_i) = \frac{1}{2}\hat{y}^jN_{ji}\ee 
and, hence, 
\be \label{E_iK}
dw_i -
\sqrt{-1}F_{ijk}N^{jl}(w_l-\Bar{w}_l)dz^k =
dw_i -\frac{1}{2}\hat{y}^jF_{ijk}dz^k 
= \frac{1}{2}(d\hat{x}_i + F_{ij}d\hat{y}^j)\, .\ee 
Therefore, the right-hand side of \re{ThmEqu'} is given by 
\be -2\sqrt{-1}N^{ij}(d\Bar{w}_j -
\sqrt{-1}\Bar{F}_{jkl}N^{km}(w_m-\Bar{w}_m)d\Bar{z}^l) = -\sqrt{-1}N^{ij}(d\hat{x}_j + 
\Bar{F}_{jk}d\hat{y}^k)\, .\ee 
Comparing this with \re{final J2dzEqu'} yields \re{ThmEqu'}. \qed 

In order to compute the para-hyper-K\"ahler metric of $N$ in the 
holomorphic coordinates $(z^i,w_i)$, it is useful to introduce 
a (pseudo-)unitary co-frame  
\be e^i = \sum e^i_Idz^I\, ,\ee 
with respect to 
the sesquilinear (pseudo-)Hermitian metric 
\be h=g+e\o = 
\sum -N_{IJ}dz^Id\Bar{z}^J\ee 
on $(M,J)$, i.e.,\   
\be  N_{IJ} = -\sum \eta_{ij}e^i_I{\Bar{e}^j}_J\, ,\quad 
(\eta_{ij}) = {\rm diag}(\mathbbm{1}_k,-\mathbbm{1}_l)\ee
where $(k,l)$ is the signature of $h$. 
We also put
\be E_i := 2\sum e_i^I(dw_I-\sqrt{-1}F_{IJK}N^{JL}(w_L-\Bar{w}_L)dz^K)\quad 
\mbox{and}\quad
E^i := \sum \eta^{ij}E_j\ee
where 
\be (e_i^I) = (e_I^i)^{-1}\quad \mbox{and}\quad (\eta^{ij}) 
= (\eta_{ij})^{-1}\, .\ee 

Now we express the full para-hyper-K\"ahler structure 
on $N$ in the co-frame $(e^i,E^i)$ of $\wedge^{1,0}_{J_1}T^*N$. 
\bt Let $(M,J,g,\n )$ be a special K\"ahler manifold and 
$(J_1, J_2, J_3, g_N)$ the para-hyper-K\"ahler structure on $N=T^*M$ 
constructed in Theorem 
\ref{c31/30Thm}. In the co-frame $(e^i,E^i)$, the 
para-hyper-K\"ahler structure has the following expression:    
\begin{enumerate}
\item[(i)]  The para-hypercomplex structure 
$(J_1,J_2,J_3=J_1J_2)$ is given by 
\bea J_1^*e^i &=& \sqrt{-1}e^i\, ,\quad J_1^*E^i = \sqrt{-1}E^i\\
     J_2^*e^i &=& \sqrt{-1}\Bar{E}^i \, ,\quad 
J_2^*E^i = \sqrt{-1}\Bar{e}^i\, .\eea  
\item[(ii)] The $J_1$-sesquilinear 
(pseudo)-Hermitian metric $h_N := g_N +e\o_1$, $\o_1=g_N(J_1\cdot, \cdot)$,  
is given by 
\be h_N = \sum \eta_{ij}(e^i\Bar{e}^j - E^i\Bar{E}^j)\, .\ee
\end{enumerate} 
\et 

\pf (i) follows from Proposition \ref{J2Prop'}.\\ 
(ii) $h_N$ is the unique $J_1$-sesquilinear form such that 
${\rm Re}\, h_N = g_N$. Therefore it is sufficient to check
that 
\be g_N = {\rm Re}  \sum \eta_{ij}(e^i\Bar{e}^j - E^i\Bar{E}^j)\, 
\label{Metrik1}.\ee
In order to check this, we calculate the right-hand side in the
affine coordinates. 
The first term is
\bea
\label{Metrik2}
g &=& 
{\rm Re} \sum e^i \Bar{e}^i \\ 
&=& - {\rm Re} \left( N_{IJ} dz^I d \overLine{z}^J \right) \nonumber \\ &=&
- \left( N_{IJ} + \ft{\der u^K}{\der x^I} N_{KL} \ft{\der u^L}{\der x^J}
\right)
dx^I dx^J  
- 2 \left( \ft{\der u^K}{\der x^I} N_{KL} \ft{\der u^L}{\der y^J}
\right) dx^I dy_J  
- \left( \ft{\der u^K}{\der y^I} N_{KL} \ft{\der u^L}{\der y^J}
\right) dy_I dy_J \nonumber 
\eea
The second term is
\bea
\label{Metrik3}
 & & 
- {\rm Re} \sum E_i \Bar{E}_i  \\ &=&
4 {\rm Re} \left( N^{IJ} ( d w_I - i F_{IKL} 
N^{KM} (w_M - \overline{w}_M) dz^L)
( d \overline{w}_J - i \overline{F}_{JNP} N^{NQ} (w_Q - \overline{w}_Q) 
d \overline{z}^P) \right)
\nonumber \\
&=&
  N^{IJ} d \hat{x}_I d \hat{x}^J 
+ N^{IK} R_{KJ} d \hat{x}_I d \hat{y}^J 
+ \ft14 ( N_{IJ} + R_{IK} N^{KL} R_{LJ} ) d \hat{y}^I d 
\hat{y}^J \;, \nonumber
\eea
where we used that \refeq{E_iK}.
Following the same steps as for the Euclidean para-$c$-map,
the metric takes the form \refeq{HKM}: 
\be
g_N  = \left( \begin{array}{cc}
g & 0 \\ 0 \;\; & - g^{-1} \\
\end{array} \right)\;.
\ee
\qed

\section{Dimensional reduction of the four-dimensional
vector multiplet \\ lagrangian}

In this section we obtain physical realizations of both
para-$c$-maps by the dimensional reduction of four-dimensional
vector multiplet lagrangians, which are reviewed in Section
3.1. In Section 3.2 and 3.3 we start with the 
bosonic part of the four-dimensional
Euclidean vector multiplet lagrangian, whose scalar target 
manifold is a special para-K\"ahler manifold $M$. After 
dimensional reduction and dualization of the gauge fields
we obtain a sigma model whose target space $N$ is 
seen to be para-hyper-K\"ahler by comparison to the results of
the previous section. This gives us a physical realization of
the Euclidean para-$c$-map $c^{4+0}_{3+0}: M \rightarrow N$.
In Section 3.4 we discuss the reduction of the four-dimensional
Minkowskian vector multiplet lagrangian over time. Here the
target space $M$ of the four-dimensional theory is
special K\"ahler,\footnote{The calculation also applies to 
special pseudo-K\"ahler manifolds, i.e., to the case of indefinite
signatures. 
}
while the target space $N$ of the 
Euclidean three-dimensional theory is again para-hyper-K\"ahler.
Thus we obtain a physical realization of the Minkowskian
para-$c$-map $c^{3+1}_{3+0}: M \rightarrow N$.

\subsection{Four-dimensional bosonic lagrangians}

It was shown in \cite{CMMS} that the general lagrangian for 
${\cal N}=2$ vector multiplets can be written in a uniform
way for both dimension $3+1$ (Minkowski space) and 
dimension $4+0$ (Euclidean space).
In the so-called new conventions, the
bosonic part of the lagrangian takes the following form:
\bea
{\cal L}^{\mscr{4+0/3+1}}  &=&
- \ft12 N_{IJ}(X,\overLine{X}) \der_m X^I \der^m \overLine{X}^J 
\nonumber \\
 & & -\ft{\Imag}{2} \left( F_{IJ}(X) F^I_{-|mn} F^{J|mn}_-
-  \overLine{F}_{IJ}(\overLine{X}) F^I_{+|mn} F^{J|mn}_+ \right) \;.
\label{4dLagrangian}
\eea
The symbol $\Imag$ represents the para-imaginary unit $e$, $e^2 =1$, 
in Euclidean signature and
the imaginary unit $i=\sqrt{-1}$ in Minkowski
signature. The space-time indices $m,n,\ldots$ 
take the values $1,\ldots, 4$ in Euclidean and
$0,\ldots, 3$ in Minkowski signature, while
the index $I=1,\ldots,n$ labels 
$n$ vector multiplets.
The scalar fields $X^I$ are (para-)complex, and
$\overLine{X}^I$ denote the (para-)complex conjugated fields.
The field strength $F^I_{mn}$ have been split into their
selfdual and antiselfdual parts with respect to the
Hodge-$\ast$-operator, according to
\be
F^I_{\pm | mn} = \ft12 \left(  F^I_{mn} \pm \ft1{\Imag}\tilde{F}^I_{mn} \right)\;.
\ee
Here $\tilde{F}^I_{mn} = \ft12 \epsilon_{mnpq} F^{I|pq}$ is the
dual field strength, and 
the convention for the $\epsilon$-tensor is
$\epsilon_{1234} =1$ for Euclidean signature and
$\epsilon_{0123} = 1$ for Minkowski signature.
Note that in Euclidean signature the (anti)selfdual field strengths
are para-complex and have eigenvalues $\pm e$ under the 
Hodge-$\ast$-operator. This non-standard definition is necessary in
order that the Euclidean lagrangian takes the same form as
the Minkowskian one.\footnote{In \cite{CMMS} we also rewrote the
Euclidean lagrangian and supersymmetry transformation rules 
in a form where all bosonic fields are real and the para-complex
unit $e$ does not appear. But then the complete analogy with 
the Minkowskian theory is lost.}

All the couplings in the lagrangian are encoded in a single
(para-)holomorphic function of the scalar fields,
the prepotential $F(X)$. We adopt the following standard definitions:
\be
F_I = \ft{\der}{\der X^I} F\;, \; \;
F_{IJ} = \ft{\der }{\der X^I} \ft{\der}{ \der X^J} F \;,\;\;
\overLine{F}_I = \ft{\der }{\der \overLine{X}^I} \overLine{F}\;, \;\;
\mbox{etc.}
\label{S1}
\ee
and
\be
N_{IJ} = \Imag \left( F_{IJ} - \overLine{F}_{IJ} \right) \;,\;\;\;
R_{IJ} = F_{IJ} + \overLine{F}_{IJ} \;.
\label{S2}
\ee

The scalar part of the lagrangian (\ref{4dLagrangian}) is a 
sigma model, i.e., the scalar fields $X^I$ can be interpreted as
the compositions of a map $X$ from space-time into a (para-)complex 
manifold $M$ with the (para-)holomorphic coordinate maps $z^I$.
{}From the lagrangian we can read off that $M$ is equipped with a
(para-)Hermitian metric\footnote{For a Minkowski signature theory 
with standard kinetic terms, $N_{IJ}$ is positive definite,
while the metric $g$ is negative definite. But all the
results derived in the following apply to arbitrary signature.}
$g = {\rm Re}\, h$, 
where  
\be
h  = - N_{IJ} dX^I d \overLine{X}^J \;.
\label{MetricM}
\ee
Moreover the relations (\ref{S1}) and (\ref{S2}) imply 
that this metric is in addition (para-)K\"ahler and in fact
special (para-)K\"ahler, because 
it has a (para-)K\"ahler potential
$K(X,\overLine{X})$,
\be
N_{IJ} = \frac{\der}{ \der X^I} \frac{\der}{ \der \overLine{X}{}^J }
K(X,\overLine{X})  \;,
\ee
which is determined by the (para-)holomorphic prepotential,
\be
K(X,\overLine{X}) = \Imag \big(F_I \overLine{X}^I - X^I \overLine{F}_I\big) \;.
\ee
As mentioned below Definition 3 in Section 2, 
the existence of a (para-)holomorphic prepotential is equivalent to 
$M$ being special (para-)K\"ahler. More details and references
can be found in \cite{CMMS}, where we also give 
the full lagrangian, including all fermionic terms and the
supersymmetry transformation rules.

\subsection{Dimensional reduction of the Euclidean lagrangian}

We will now perform the dimensional reduction of the 
Euclidean lagrangian ${\cal L}^{4+0}$. The 
vector potential $A^I_m$ of the four-dimensional
field strength $F^I_{mn} = \der_m A^I_n - \der_n A^I_m$
decomposes into a three dimensional vector
$A^I_a$, $a=1,2,3$ and a scalar $p^I$:
\be
(A^I_m) = (A^I_a, p^I :=  A^I_4 = A^{I|4}) \;.
\ee
The resulting three-dimensional lagrangian is 
\bea
{\cal L}^{4+0}_{3+0}  &=&
- \ft12 N_{IJ} \der_a X^I \der_a X^J 
- \ft12 N_{IJ} \der_a p^I \der_a p^J \nonumber \\
 & & + \ft12 R_{IJ} \der_a p^I \epsilon_{abc} F_{bc}^J 
 - \ft14 N_{IJ} F^I_{ab} F^J_{ab}  \;,
\eea
where
$F^I_{ab} = \der_a A^I_b - \der_b A^I_a$. The three-dimensional
$\epsilon$-tensor $\epsilon_{abc}$ is normalized 
such that $\epsilon_{123} = 1$.

In three dimensions we can dualize the $n$ abelian vector fields
$A^I_a$ into $n$ scalar fields $s_I$, so that we obtain
a dual lagrangian which depends on $4n$ real scalar fields.
We first introduce the dual
gauge fields
\be
H^I_a := \ft12 \epsilon_{abc} F^I_{bc} \Leftrightarrow 
F^I_{ab} =  \epsilon_{abc} H^I_c \,.
\label{DualGF}
\ee
Then we promote the Bianchi identity $\epsilon_{abc} \der_a F^I_{bc}=0$
of the field
strength $F^I_{ab}$ to a field equation, using
Lagrange multiplier fields $s_I$:
\be
\hat{\cal L}^{4+0}_{3+0} = {\cal L}^{4+0}_{3+0} 
+  \der_a s_I H_a^I \;.
\label{LhatEuclid}
\ee
As usual, the field equation for $s_I$ is the Bianchi
identity for $F^I_{ab}$, which can be solved
by introducing the vector potential $A^I_a$. 
Plugging this back into $\hat{\cal L}^{4+0}_{3+0}$
we recover ${\cal L}^{4+0}_{3+0}$. But instead
we can first solve the equation of motion for $H^I_a$, which
is 
\be
H_a^I = N^{IJ} \der_a s_J +
N^{IJ} R_{JK} \der_a p^K \;,
\ee
where $N^{IJ}$ is the inverse of the matrix $N_{IJ}$. 
Plugging this into $\hat{\cal L}^{4+0}_{3+0}$
we obtain the dual lagrangian
\bea
\label{DualIIb}
\tilde{\cal L}^{4+0}_{3+0}  &=& - \ft12 N_{IJ} \der_a X^I \der_a 
\overLine{X}^J
- \ft12 N_{IJ} \der_a p^I \der_a p^J \nonumber \\
 && + \ft12  R_{IK} N^{KL} R_{LJ} \der_a p^I \der_a p^J 
+ \ft12  N^{IJ} \der_a s_I \der_a s_J  \nonumber \\
&&
+  N^{IK} R_{KJ} \der_a s_I \der_a p^J \;.
\eea
The lagrangians 
${\cal L}^{4+0}_{3+0}$ and 
$\tilde{\cal L}^{4+0}_{3+0}$ are dual in the sense that they
give rise to equivalent field equations despite that they have a different
field content. Hence, they are interpreted as two different
lagrangian descriptions of the same theory. Note that they are not related by
a local field redefinition. Thus the duality is not a symmetry of a given
lagrangian.

\subsection{The para-hyper-K\"ahler geometry of the reduced lagrangian}

The dual lagrangian depends on $4n$ real scalar fields 
$\mbox{Re}(X^I), \mbox{Im}(X^I), p^I, s_I$. We denote the 
resulting target space by $N$, and our goal is to prove that
$N$ is (for given $M$) isometric to the para-hyper-K\"ahler
manifold $N$ constructed in Section 2. 

Since half of the real fields, namely those parametrizing $M$,  
combine into the 
para-complex fields $X^I$, it is natural to wonder whether one
can also combine the real fields $p^I,s_I$, which were obtained by dimensional
reduction, into para-complex fields, such that the metric of $N$
is para-Hermitian.  This is indeed possible: defining $n$
para-complex fields $W_I$ by 
\be
W_I = \ft12 \left( s_I + R_{IJ} p^J + e N_{IJ} p^J \right) 
= \ft12 s_I +  F_{IJ} p^J \;,
\label{Wfields}
\ee
the lagrangian (\ref{DualIIb}) takes the form 
\bea
\tilde {\cal L}^{4+0}_{3+0} &=& 
- \ft12 N_{IJ} \der_a X^I \der_a \overLine{X}^J \nonumber\\
&+& 2 N^{IJ} \left( \der_a W_I - e F_{IKP} N^{KM} (W_M - \overLine{W}_M)
\der_a X^P \right)   \nonumber \\
 & & \cdot \left( \der_a \overLine{W}_J - e \overLine{F}_{JLQ} N^{LN}
(W_N - \overLine{W}_N) \der_a \overLine{X}^Q ) \right) \;.
\label{DualIIbHerm}
\eea
{}From this one reads off the sesquilinear para-Hermitian metric $h_N'$ of $N$:
\bea
h_N' & = &  - N_{IJ} dX^I d \overLine{X}^J 
+ 4 N^{IJ}  \left( dW_I  - e F_{IKP} N^{KM} (W_M - \overLine{W}_M)
d X^P \right) \nonumber \\
 && 
\cdot 
\left( d \overLine{W}_J - e \overLine{F}_{JLQ} N^{LN} ( W_N - \overLine{W}_N)
d \overLine{X}^Q \right) \;.
\label{MetricIIb}
\eea
Given (\ref{Wfields}), the verification of (\ref{DualIIbHerm}) is
straightforward, though somewhat tedious. 
When performing the calculation, it is useful to note the
identity
\be
N^{IJ} F_{IK} \overline{F}_{JL} S^{KL} =
\ft14 \left( N^{IJ} R_{IK} R_{JL} - N_{KL} \right) S^{KL} \;,
\ee
which holds for any symmetric matrix $S^{KL}$.

Since the choice
(\ref{Wfields}) is not so obvious, let us add the following 
remarks. The idea is to find para-complex fields $W_I$ such that 
the resulting metric is manifestly para-Hermitian, i.e.,
the purely (anti-)para-holomorphic components vanish identically.
Once we decide that $s_I$ is contained in the real part of $W_I$, 
we know that 
$p^J$ must be present in the imaginary part, and the coefficient must
be adjusted in such a way that all purely (anti-)para-holomorphic
terms cancel. It is easy to see that
this coefficient must depend on $X^I$, and the obvious candidates 
are $F_{IJ}$ and its real and imaginary part. 

The fields $X^I, W_I$ can be viewed 
as compositions of maps from space-time to $N$ with corresponding 
para-holomorphic coordinate maps $z^I,w_I$. 
In comparison to Section 2, the para-holomorphic coordinates $(z^I,w_I)$ 
define an identification of the target space $N$ with the para-holomorphic
cotangent bundle $\wedge^{1,0}T^*M\cong T^*M$. This 
identification\footnote{\label{Fussnote} Equivalently, we could have used the
real coordinates associated with the scalar fields $({\rm Re}\, X^I, 
{\rm Re}\, F_I, s_I, 2p^I)$ to identify the target manifold $N$ with
the cotangent bundle $T^*M$. This identification yields the same 
para-complex structure $J_1'$ on $N$, namely, the canonical one induced 
by the para-complex structure $J$ on $M$. The factor 2 in front of $p^I$ 
is chosen such that the above scalar fields correspond to the
coordinates $(x^I,y_I,\hat{x}_I,\hat{y}^I)$ of Section 2, cf.\ \re{wIdzIEqu} 
and \re{Wfields}.}  
defines  a para-complex 
structure $J'_1$ on $N$  for which the coordinates $(z^I,w_I)$ 
are para-holomorphic.  Moreover we 
have found a metric $g_N'={\rm Re}\, h_N'$ which is para-Hermitian 
with respect 
to $J'_1$. In fact it is easy to see that $g_N'$ is para-K\"ahler
with respect to $J'_1$, because 
\be
K =  - e (F_I \overLine{X}^I - \overLine{F}_I X^I ) 
- 2 N^{IJ} (W_I - \overLine{W}_I) (W_J - \overLine{W}_J)
\ee
is a $J'_1$-para-K\"ahler potential:
\be
h_N' = \der \overLine{\der} K \;,
\ee
where
\be
\der = dX^I \frac{\der}{\der X^I} + d W_I \frac{\der}{\der W_I} 
\;\;\;\mbox{and} \;\;\;
\overLine{\der} = d\overLine{X}^I \frac{\der}{\der \overLine{X}^I} + d 
\overLine{W}_I \frac{\der}{\der \overLine{W}_I} 
\;.
\ee
In order to show that $g_N'$ is even para-hyper-K\"ahler we 
would have to proceed as follows:
first, find a second para-complex structure $J'_2$ which 
anticommutes with $J'_1$, so that $J'_3 = J'_1 J'_2$ is a complex
structure. This 
gives us an almost para-hypercomplex structure (see Definition 4 in
Section 2). Second, verify that $J'_1, J'_2$ are
anti-isometries of $g_N'$, while $J'_3$ is an isometry,
so that we have an almost para-Hermitian-structure.
Finally, check that the three corresponding 
fundamental two-forms are closed. By a variant of the so-called
Hitchin lemma this implies that the $J'_\alpha$ are integrable
and parallel, so that $g_N'$ is a para-hyper-K\"ahler metric.
Note that as the starting point for all these calculations we would have
to make an educated guess for $J'_2$, first. But now we can 
profit enormously from the results of Section 2. 
By comparing (\ref{MetricIIb}) to (\ref{dsmReal0}),(\ref{dsmReal}) 
and (\ref{dsmReal2}) we see that metric obtained by dimensional
reduction of the lagrangian takes the form
\be
g'_N  = {\rm Re} h'_N = {\rm Re} \sum ( e^i \overline{e}^i -
E_i \overline{E}_i )
\ee 
where $e^i$ and $E_i$ are given by (\ref{ei}), (\ref{NIJ})  and (\ref{Ei}), 
respectively. Now we can use Theorem 4 in combination with Remark 1. 
Let $(M,J,g,\nabla)$ be the special para-K\"ahler manifold underlying
the four-dimensional lagrangian. Then $(M,J,g,\nabla')$ with 
$\nabla' = J \circ \nabla \circ J^{-1}$ is also a special para-K\"ahler 
manifold, to which we can associate, by Theorem 4, a 
para-hyper-K\"ahler manifold $(N, J'_\alpha, g_N)$ with metric
$g_N = g \oplus g^{-1} = 
{\rm Re} \sum ( e^i \overline{e}^i + E'_i \overline{E}'_i)$,
where the decomposition of $TN$ is defined by $\nabla'$. 
By the diffeomorphism $(X^I, W_J) \rightarrow (X^I, e W_J)$ of
Remark 1, this manifold is mapped to the para-hyper-K\"ahler 
manifold $(N,J_\alpha, g'_N)$, with metric
$g'_N = g \oplus (-g)^{-1} = 
{\rm Re} \sum ( e^i \overline{e}^i - E_i \overline{E}_i)$,
where the decomposition of $TN$ is defined by $\nabla$. 
According to (\ref{MetricIIb}) this is the para-hyper-K\"ahler 
manifold which is associated, through dimensional reduction 
of the four-dimensional lagrangian, to the special para-K\"ahler
manifold $(M,J,\nabla)$. 
In particular, the metric (\ref{MetricIIb}) is para-hyper-K\"ahler
and the (para-)complex structures $J_\alpha$ and (para-)K\"ahler 
forms $\omega_\alpha$ 
can be read off from the formulae derived in 
Section 2. Moreover, we see that the dimensional reduction of lagrangians
provides a physical realization of the Euclidean para-$c$-map
$c^{4+0}_{3+0}$ of Section 2.


\subsection{Reduction of the Minkowski lagrangian over time}

The second way to define a three-dimensional theory with
signature $3+0$ is to reduce the Minkowskian version
of (\ref{4dLagrangian}) over time. Since this is very similar
to the reduction we discussed above, we only mention some
key formulae.\footnote{We also refer to \cite{CMMS} for
a detailed comparision between the dimensional reduction
over time compared to the dimensional reduction over space.} 
The decomposition of the four-dimensional 
gauge field is
\be
(A^I_m) = (A^I_a, p^I := - A^I_0 = A^{I|0}) \;, \;\;\; a=1,2,3 \;.
\ee
and the reduced lagrangian is
\bea
{\cal L}^{3+1}_{3+0}  &=&
- \ft12 N_{IJ} \der_a X^I \der_a X^J 
+ \ft12 N_{IJ} \der_a p^I \der_a p^J \nonumber \\
 & & - \ft12 R_{IJ} \der_a p^I \epsilon_{abc} F_{bc}^J 
 - \ft14 N_{IJ} F^I_{ab} F^J_{ab}  \;.
\eea
As in Section 3.2, we introduce dual gauge fields
by (\ref{DualGF}) and promote the Bianchi identity to
a field equation using a Lagrange multiplier:
\be
\hat{\cal L}^{3+1}_{3+0} = {\cal L}^{3+1}_{3+0} - \der_a s_I H^I_a \;.
\label{LhatMink}
\ee
Note that we took a different relative sign on the right hand side
compared to (\ref{LhatEuclid}). This has been done for later convenience.
The prefactor of this term does not have an intrinsic 
meaning, as it can be compensated by rescaling the field $s_I$. 

Integrating out the gauge fields $F^I_{ab}$ we obtain the dual
lagrangian
\begin{eqnarray}
  \tilde{\cal L}^{3+1}_{3+0}  &=& - \ft12 N_{IJ} \der_a X^I \der_a 
  \overLine{X}^J
  + \ft12 N_{IJ} \der_a p^I \der_a p^J \nonumber \\
  && + \ft12  R_{IK} N^{KL} R_{LJ} \der_a p^I \der_a p^J 
  + \ft12  N^{IJ} \der_a s_I \der_a s_J  \nonumber \\
  &&
  + N^{IK} R_{KJ} \der_a s_I \der_a p^J  \;.
\label{DualIIa}
\end{eqnarray}
The structure of this lagrangian is similar to the one of 
(\ref{DualIIb}) but differs in its distribution 
of relative signs. Since $M$ is now complex rather than para-complex,
while $p^I$ comes from the time-like component of a gauge field,
the kinetic terms of both $\mbox{Re}(X^I)$ and $\mbox{Im}(X^I)$
have the same sign, while $p^I$ and $s_I$ come with 
the opposite sign.\footnote{
Here we assume that $N_{IJ}$ is positive definite. The generalization to
indefinite signature is straightforward.}

As already in the original sigma model with target space $M$,
$2n$ of the real scalars combine into 
the $n$ complex scalar fields $X^I$. The other
$2n$ real scalars $p^I, s_I$ can be combined into $n$ complex
scalar fields by
\be
W_I = \ft12 \left( s_I + R_{IJ} p^J + i N_{IJ} p^J \right) 
= \ft12 s_I +  F_{IJ} p^J \;,
\ee
and by rewriting 
the dual lagrangian (\ref{DualIIa}) in terms of these fields 
we obtain 
\begin{eqnarray} \label{lagrEqu} 
  \tilde {\cal L}^{3+1}_{3+0} &=& 
  - \ft12 N_{IJ} \der_a X^I \der_a \overLine{X}^J \nonumber\\
  &+& 2 N^{IJ} \left( \der_a W_I - i F_{IKP} N^{KM} (W_M - \overLine{W}_M)
    \der_a X^P \right)  \nonumber \\
  & & \cdot \left( \der_a \overLine{W}_J - i \overLine{F}_{JLQ} N^{LN}
    (W_N - \overLine{W}_N) \der_a \overLine{X}^Q ) \right) \;.
\end{eqnarray}
The corresponding metric $g_N$ on $N$ is readily seen to coincide with 
the metric specified by (\ref{Metrik1}), (\ref{Metrik2}), (\ref{Metrik3}), 
which we obtained in Section 2.7 by applying the Minkowskian 
para-$c$-map $c^{3+1}_{3+0}$ to the special K\"ahler manifold $M$.
In particular, $g_N$ is para-hyper-K\"ahler.

Note that the para-hyper-K\"ahler metrics which can be obtained
by the para-$c$-maps are not the most general para-hyper-K\"ahler 
metrics, but only a subset. This is clear from the large number of
isometries. In particular,  constant real shifts of the 
fields $W_I$ obviously preserve the lagrangians \re{DualIIbHerm},\re{lagrEqu}.
More generally, looking at \re{DualIIb}, \re{DualIIa},  we see that constant 
shifts of the fields $p^I, s_I$  are manifest symmetries of 
the lagrangian. Geometrically this corresponds to the fact that 
translations in the fiber coordinates $(\hat{x}_I,\hat{y}^I)$ are
isometries of the metric \re{c4030metricEqu}, \re{HKM}, cf.\ footnote \ref{Fussnote}. Physically, 
these isometries
correspond to the gauge symmetries of the gauge fields $A^I_m$ which 
have been transformed into the scalars $p^I,s_I$ by dimensional reduction
and dualization. Since it is well known that for Minkowskian
hypermultiplets
{\em any} hyper-K\"ahler manifold is an admissible target space, we conjecture
that {\em any} para-hyper-K\"ahler manifold is an
admissible target space for Euclidean hypermultiplets. 
The analogous result for vector multiplets
was proven in \cite{CMMS}, and we expect that our conjecture can be
proven by similar methods. Also note that the scalar geometry of a 
hypermultiplet is inert under dimensional reduction, because it does not
contain bosonic fields other than scalars. Therefore we can lift
our lagrangian to $(4+0)$ dimensions. We expect that this extends to 
the fermionic part of the lagrangian in the same way as it works
for Minkowskian hypermultiplets \cite{DDKV}. Moreover one should
be able to dimensionally lift the action to higher-dimensional Euclidean 
supersymmetric actions up to the point where no Euclidean supersymmetry algebra
with eight real supercharges exists.

One should expect that the dimensional reduction of four-dimensional
vector multiplets does not give the most general hypermultiplet manifolds.
The reason is that we used a `classical' dimensional reduction where 
we ignored all the massive Kaluza-Klein states. In an exact treatment
one would have to integrate them out, which results in modified 
couplings between the lower-dimensional massless fields. Put differently,
the metric which we computed is the classical approximation of 
the full metric, which also receives perturbative threshold corrections from
integrating out massive fields.
Moreover, there will also be non-perturbative corrections due to 
field configurations with finite action, which wind around the compact
direction. Such instantons are expected to break the continuous
shift symmetries which we discussed above to a discrete subgroup.
Therefore we expect that after the inclusion of these corrections, 
the hypermultiplet manifolds
are more generic than those constructed here. The investigation
of such manifolds will be postponed to future work.

Following the terminology used for hyper-K\"ahler manifolds we will call
those para-hyper-K\"ahler manifolds which can be obtained by one of the
para-$c$-maps {\em special para-hyper-K\"ahler manifolds}. There is no
reason to believe 
that if a para-hyper-K\"ahler manifold can be constructed by one of
the para-$c$-maps, it can also be constructed by the other.
While we have a lot of freedom 
in choosing $M$ (we can pick any
holomorphic or para-holomorphic function, respectively),
the construction of $N=T^*M$ is then completely fixed. However,
there is a subclass of the special para-hyper-K\"ahler 
manifolds which can be obtained by both para-$c$-maps. The reason 
is that we can start with a five-dimensional vector multiplet 
lagrangian and first reduce over time and then over space,
or vice versa. As the result of the dimensional reduction should not
depend on the order of steps, this gives us the desired subclass
of {\em very special para-hyper-K\"ahler manifolds}. This will be the
subject of Section 4.

In \cite{CMMS} we showed that the full vector multiplet lagrangian
can be written in a uniform way, which applies to both Minkowski and
Euclidean signature. Here we note a similar result for the bosonic
part of hypermultiplet lagrangians. Indeed, 
using the symbol $\ih$, the lagrangians  
$\tilde {\cal L}^{4+0}_{3+0}$  and
$\tilde {\cal L}^{3+1}_{3+0}$ take the following form:\footnote{At this
point the different signs in (\ref{LhatEuclid}) and (\ref{LhatMink})
turn out to be convenient.}
\begin{eqnarray}
  \tilde {\cal L}^{4+0,\,3+1}_{3+0,\,3+0} &=&
  - \ft12 N_{IJ} \der_a X^I \der_a \overLine{X}^J \nonumber\\
  &+& 2 N^{IJ} \left( \der_a W_I - \ih F_{IKP} N^{KM} (W_M - \overLine{W}_M)
    \der_a X^P \right)  \nonumber \\
  & & \cdot \left( \der_a \overLine{W}_J - \ih \overLine{F}_{JLQ} N^{LN}
    (W_N - \overLine{W}_N) \der_a \overLine{X}^Q ) \right)  \;.
\end{eqnarray}

The corresponding sigma model metric $g_N={\rm Re}\, h_N$ is the real part of 
\begin{eqnarray}
  h_N &=&  -N_{IJ} dX^I d \overLine{X}^J 
  + 4 N^{IJ}  \left( dW_I  - \ih F_{IKP} N^{KM} (W_M - \overLine{W}_M)
    d X^P \right) \nonumber \\
  && 
  \cdot 
\left( d \overLine{W}_J - \ih \overLine{F}_{JLQ} N^{LN} ( W_N - \overLine{W}_N)
    d \overLine{X}^Q \right)  \;.
  \label{MetricIIa}
\end{eqnarray}
The (para-)K\"ahler potential of $g_N$ with respect to the 
(para-)complex structure $J_1$ is
\begin{equation}
K =  -\ih (F_I \overLine{X}^I - \overLine{F}_I X^I ) 
- 2 N^{IJ} (W_I - \overLine{W}_I) (W_J - \overLine{W}_J)\;,
\end{equation}
with
\begin{equation}
  h_N = \der \overLine{\der} K
\end{equation}
where
\begin{equation}
\der = dX^I \frac{\der}{\der X^I} + d W_I \frac{\der}{\der W_I} 
\;\;\;\mbox{and} \;\;\;
\overLine{\der} = d\overLine{X}^I \frac{\der}{\der \overLine{X}^I} + d 
\overLine{W}_I \frac{\der}{\der \overLine{W}_I} 
\;.
\end{equation}
This reflects that in the framework of complex-Riemannian geometry
hyper-K\"ahler and para-hyper-K\"ahler geometry can be interpreted 
as real forms of the same complex geometry. We expect that this
will be useful when extending the above result to the full
hypermultiplet lagrangian.

\section{Dimensional reduction of the five-dimensional vector 
multiplet \\ lagrangian \label{5to3}} 

In \cite{CMMS} we discussed the dimensional
reduction of the general lagrangian for $(4+1)$-dimensional
vector multiplets with respect to both time and space.
The resulting $(4+0)$-dimensional and $(3+1)$-dimensional
lagrangians are of the type 
(\ref{4dLagrangian}). In both cases one can perform
a further reduction to $3+0$ dimensions,
and one expects that the resulting lagrangians
are equivalent. We will now verify this for the bosonic
parts of the lagrangians. 

The bosonic fields of the five-dimensional lagrangian
are $n$ real scalars $\sigma^I$ and $n$ gauge fields
$A^I_\mu$, where $\mu =0,\ldots,4$. All couplings are encoded
in a real prepotential ${\cal V}(\sigma)$ which is a general 
cubic polynomial in the fields $\sigma^I$. The two following
expressions appear explicitly in the lagrangian:
\be
a_{IJ} = a_{IJ}(\sigma) = \ft{\der}{\der \sigma^I}\ft{\der}{\der \sigma^J}
{\cal V}(\sigma)
\ee
and
\be
d_{IJK} = \ft{\der}{\der \sigma^I}\ft{\der}{\der \sigma^J}
\ft{\der}{\der \sigma^K} {\cal V}(\sigma) = \mbox{const.}
\ee
The symmetric matrix $a_{IJ}(\sigma)$ can be interpreted as
the metric of a {\em very special real manifold} $K$,\footnote{By 
definition, these are real manifolds with a metric defined by a real
cubic polynomial, see \cite{CMMS}.}
which is parametrized by the $n$ real scalars $\sigma^I$.

If we first
reduce with respect to space, then one spatial component
of each gauge field becomes a scalar, $b^I \sim A^I_4$,
and combines with $\sigma^I$  into a complex scalar $X^I=\sigma^I + i b^I$.
The bosonic lagrangian is the Minkowskian version of
(\ref{4dLagrangian}) with a holomorphic prepotential $F(X)$, which
is determined by the real prepotential of the five-dimensional
theory through
\be
\left. F(X) \right|_{b^I =0}  = \frac{1}{2i} {\cal V}(\sigma) \;.
\ee
Thus the prepotential $F(X)$ is cubic with purely imaginary coefficients.
The couplings take the special form
\be
R_{IJ} = R_{IJ}(b) = d_{IJK} b^K \;,\;\;\;N_{IJ} = a_{IJ} (\sigma)  \;.
\ee
The scalar manifolds $M$ obtained this way are 
called {\em very special
K\"ahler manifolds}, and $r^{4+1}_{3+1}: K \rightarrow M$ is
called the $r$-map.

If we further reduce this model with respect to time, then
the time-like components of the gauge fields become scalars, 
$p^I \sim A^I_0$. 
The reduced lagrangian takes the form \\
\bea
{\cal L}^{3+1\,, \;\;4+1}_{3+0\,, \;\;3+1}  &=&
- \ft12 a_{IJ} \der_a \sigma^I \der_a \sigma^J
- \ft12 a_{IJ} \der_a b^I \der_a b^J
+ \ft12 a_{IJ} \der_a p^I \der_a p^J \nonumber \\
 & & - \ft12 d_{IJK} b^K \der_a p^I \epsilon_{abc} F_{bc}^J 
 - \ft14 a_{IJ} F^I_{ab} F^J_{ab} \;.
\label{VerySpecialIIa}
\eea
Dualizing the gauge fields into scalars we obtain a 
para-hyper-K\"ahler manifold $N$, which is determined by
a real cubic prepotential ${\cal V}(\sigma)$. We call
manifolds, which are obtained by the
successive $r$-map and para-$c$-map, 
{\em very special para-hyper-K\"ahler manifolds}.

Let us now consider what happens if we perform the dimensional
reductions in opposite order. We first reduce over time,
and therefore the scalars $b^I$ correspond to the time-like
components of the gauge fields, $b^I \sim A^I_0$. 
As a consequence they combine with the $\sigma^I$ into
para-complex scalars $X^I = \sigma^I + e b^I$. The 
four-dimensional lagrangian is determined by a 
para-holomorphic prepotential which is cubic with 
purely para-imaginary coefficients:
\be
 F(X) \big|_{b^I =0}  = \frac{1}{2e} {\cal V}(\sigma) \;.
\ee
The corresponding manifolds $M$ are called {\em very special
para-K\"ahler manifolds}. The map $r^{4+1}_{4+0}: K \rightarrow M$ is 
called the para-$r$-map.

If we now reduce over space, 
one space-like component of each gauge field becomes a scalar,
$p^I \sim A^I_4$. The reduced lagrangian is
\bea
{\cal L}^{4+0\,, \;\;4+1}_{3+0 \,, \;\; 4+0}  &=&
- \ft12 a_{IJ} \der_a \sigma^I \der_a \sigma^J 
+ \ft12 a_{IJ} \der_a b^I \der_a b^J 
- \ft12 a_{IJ} \der_a p^I \der_a p^J \nonumber \\
 & & + \ft12 d_{IJK} b^K \der_a p^I \epsilon_{abc} F_{bc}^J 
 - \ft14 a_{IJ} F^I_{ab} F^J_{ab}  \;.
\label{VerySpecialIIb}
\eea
{}From the relations between the scalar fields $b^I, p^I$ and
the five-dimensional gauge fields it is clear that 
that the lagrangians (\ref{VerySpecialIIa}) and (\ref{VerySpecialIIb})
must be related by $b^I \leftrightarrow p^I$. To verify this we
relabel the fields as indicated and perform a 
partial integration of the first term in the second line. Using 
the Bianchi identity $\epsilon_{abc} \der_a F_{bc}^I =0$
together with the fact that $d_{IJK}$ is constant, we see
that the lagrangians (\ref{VerySpecialIIa}) and
(\ref{VerySpecialIIb}) agree up to a total derivative and therefore
have the same equations of motion.\footnote{ 
The calculation also illustrates that this will {\em not}
generalize to non-cubic prepotentials, because in this case
$d_{IJK}$ is no longer constant.}  By dualizing the lagrangians
${\cal L}^{4+0, \; 4+1}_{3+0 ,\; 4+0}$ and 
${\cal L}^{4+0, \; 3+1}_{3+0 ,\; 3+1}$ we can find the relation between
the associated non-linear sigma models 
$\tilde {\cal L}^{4+0, \; 4+1}_{3+0 ,\; 4+0}$ and
$\tilde{\cal L}^{4+0, \; 3+1}_{3+0 ,\; 3+1}$. 
The fields of the two lagrangians are related by 
\be 
(\sigma^I, b^I, s_I, p^I) \rightarrow
(\sigma^I, \pm p^I, s_I + d_{IJK} b^J p^K, \mp b^I) \;,
\ee
and this defines an isometry between the corresponding 
very special para-hyper-K\"ahler
manifolds.
This can be summarized by 
\be
c^{4+0}_{3+0}\circ r^{4+1}_{4+0} =
q^{4+1}_{3+0} \cong c^{3+1}_{3+0}\circ r^{4+1}_{3+1}\;,
\ee
which proves Proposition 4 of Section 2. The corresponding commutative
diagram underlies Figure \ref{Diagram} in Section 1. 
    
\subsubsection*{Acknowledgements} 
This work has been supported by DFG within the `Schwerpunktprogramm
Stringtheorie' (SPP 1096). T.M. thanks the Institut de Math\'ematique
\'Elie Cartan, Universit\'e Henri Poincar\'e, for hospitality and
support during the last stages of this work.

\end{document}